\documentclass[fleqn,usenatbib]{mnras}
\usepackage[T1]{fontenc}

\DeclareRobustCommand{\VAN}[3]{#2}
\let\VANthebibliography\thebibliography
\def\thebibliography{\DeclareRobustCommand{\VAN}[3]{##3}\VANthebibliography}

\usepackage{graphicx}	
\usepackage{amsmath}	
\usepackage{amssymb}	

\usepackage{xcolor}

\usepackage{newtxtext,newtxmath}

\definecolor{c-enlace}{HTML}{2854d7}
\graphicspath{{figures/}}

\title[Thermal forces on eccentric or inclined planets]{Evolution of the eccentricity and inclination of low-mass planets subjected to thermal forces: a numerical study}

\author[Cornejo et al.]{
Sonia Cornejo,$^{1}$\thanks{E-mail: soniac@icf.unam.mx}
Fr{\'e}d{\'e}ric S. Masset,$^{1, 2}$
Raúl O. Chametla,$^{3, 1}$
and S{\'e}bastien Fromenteau$^{1}$
\\
$^{1}$Instituto de Ciencias F{\'i}sicas, Universidad Nacional Aut{\'o}noma de M{\'e}xico, Av. Universidad s/n, 62210 Cuernavaca, Mor., M{\'e}xico\\
$^{2}$University Nice-Sophia Antipolis, CNRS, Observatoire de la C{\^o}te d’Azur, Laboratoire LAGRANGE, CS 34229. 06304 Nice Cedex 4, France\\
$^{3}$Charles University, Faculty of Mathematics and Physics, Astronomical Institute, V Holešovičkách 747/2, 180 00 Prague 8, Czech Republic
}

\date{Accepted XXX. Received YYY; in original form ZZZ}

\pubyear{2022}

\begin{document}
\label{firstpage}
\pagerange{\pageref{firstpage}--\pageref{lastpage}}
\maketitle

\begin{abstract}
  By means of three dimensional, high resolution hydrodynamical simulations we study the orbital evolution of weakly eccentric or inclined low-mass protoplanets embedded in gaseous discs subject to thermal diffusion. We consider both non-luminous planets, and planets that also experience the radiative feedback from their own luminosity. We compare our results to previous analytical work, and find that thermal forces (the contribution to the disc's force arising from thermal effects) match those predicted by linear theory within $\sim 20$~\%. When the planet's luminosity exceeds a threshold found to be within $10$~\% of that predicted by linear theory, its eccentricity and inclination grow exponentially, whereas these quantities undergo a strong damping below this threshold. In this regime of low luminosity indeed, thermal diffusion cools the surroundings of the planet and allows gas to accumulate in its vicinity. It is the dynamics of this gas excess that contributes to damp eccentricity and inclination. The damping rates obtained can be up to $h^{-1}$ times larger than those due to the resonant interaction with the disc, where $h$ is the disc's aspect ratio. This suggests that models that incorporate planet-disc interactions using well-known formulae based on resonant wave-launching to describe the evolution of eccentricity and inclination underestimate the damping action of the disc on the eccentricity and inclination of low-mass planets by an order of magnitude.
\end{abstract}




\section{Introduction}
\label{sec:introduction}
Planetary embryos gravitationally interact with the protoplanetary discs out of which they form. This interaction changes their orbital elements, leading to a variation of the semi-major axis $a$ (a process known as planetary migration), and in general to a decay of the eccentricity $e$ and inclination $i$ \citep{tanaka2002,2004ApJ...602..388T}.

Previous work that has addressed this problem generally considers  planetary embryos embedded in isothermal (or adiabatic\footnote{The equations that describe the time-evolution of the orbital elements of a low-mass protoplanet in an adiabatic disc can simply be obtained from those of isothermal discs by substituting the isothermal speed of sound by the adiabatic one.}) discs. However, a more realistic view of the problem must consider that discs have finite thermal diffusivity, as they undergo radiative heat transport.  While the disc's thermal diffusivity has been taken into account in the non-linear dynamics of the corotation torque more than a decade ago \citep{2010ApJ...723.1393M,pbk11}, its more immediate effect, present in the linearised equations of the flow, has only been considered more recently. Compared to the adiabatic case, thermal diffusion enables a flow of heat away from the hot region surrounding the embryo, making the gas in the vicinity of the planetary embryo cooler and denser. Compared to the force exerted by an adiabatic disc,  there is a new contribution arising from these cold and dense surroundings
called the ``cold finger effect'' by  \citet{2014MNRAS.440..683L} and cold thermal force by \citet{2017MNRAS.472.4204M}. Furthermore, if the embryo, heated by accretion, injects energy into the surrounding gas, another component of the thermal force appears. This component of the force is named ``heating''  force since the embryo releases heat into the  gas, which becomes hotter and less dense than in the case of a non-luminous embryo \citep{2015Natur.520...63B,2017MNRAS.472.4204M}. The heating force scales with the luminosity of the protoplanet, as long as it is not too large.

From the above, it can be seen that the net effect of the thermal force depends on the planet's luminosity. When the planet has a vanishing or small luminosity, the surrounding gas, as a consequence of thermal diffusion, is colder and denser than in the adiabatic case. The planet's luminosity is said to be sub-critical.
As the luminosity increases, the surroundings become progressively hotter (and less dense), up to a point where they are actually hotter and less dense than in the adiabatic case. Beyond this critical luminosity, the thermal force reverses, and the luminosity is said to be super-critical. Namely:
\begin{enumerate}
\item For a planet on a circular orbit, the thermal force induces an inward migration if the luminosity is sub-critical \citep{2014MNRAS.440..683L} and an outward migration otherwise \citep{2015Natur.520...63B};
\item For a moderately eccentric (or inclined) planet, thermal forces induce a damping of the eccentricity (or inclination) if the luminosity is sub-critical and an excitation of the eccentricity (or inclination) otherwise \citep{2019MNRAS.485.5035F};
\item For a planet travelling at larger speed across the gas, the thermal force is aligned with its velocity vector relative to the gas, and is opposed to the direction of motion (so it is a drag force) if the luminosity is sub-critical \citep{2019MNRAS.483.4383V} while it has same direction as the motion (so it is a thrust) otherwise. \citep{2017MNRAS.465.3175M,2017arXiv170401931E,2019MNRAS.483.4383V}.
\end{enumerate}

In general, in the planet forming regions of protoplanetary discs, thermal forces on low-mass planets can be substantially larger than the forces arising from the interaction with the disc at  Lindblad's and corotation resonances \citep{2017MNRAS.472.4204M}. They therefore essentially determine the orbital evolution of these objects, and it is crucial to assess in detail their properties. Thermal forces have been studied in the three regimes outlined above using linear perturbation theory, and the analytic predictions have been subsequently confirmed by numerical simulations in the case of a planet on circular orbit \citep{2020ApJ...902...50H,2021MNRAS.501...24C} and in the case of a planet travelling at larger speed across the disc, i.e. in a context of dynamical friction \citep{2019MNRAS.483.4383V,2020MNRAS.495.2063V}. The intermediate situation, that of a planet with moderate eccentricity or inclination, has so far not been assessed by means of numerical simulations, and the analytical expressions of \citet{2019MNRAS.485.5035F} await confirmation.

This purpose of this paper is to corroborate these analytical expressions numerically, when the assumptions underlying the derivation of \citet{2019MNRAS.485.5035F} are verified, and to explore situations in which they are not. It is organised as follows. In section~\ref{sec:resumen-ecs}, we summarise the results on thermal forces obtained by linear theory. We present our setups in section~\ref{sec:setup}. In section~\ref{sec:forces}, we study both the cold and heating thermal force and compare our results to analytic expectations. In section~\ref{sec:orbitalelements}, we study the evolution of the orbital elements of the planet, allowed to move freely in the disc. Then in section~\ref{sec:discusion}, we discuss further aspects of thermal forces on eccentric and inclined low-mass protoplanets, and we summarise our results in section~\ref{sec:conclusiones}.

\section{Summary of results from linear theory}
\label{sec:resumen-ecs}
A key quantity that features in the linear theory of thermal forces onto planets on circular or moderately eccentric or inclined orbits is the thermal lengthscale $\lambda$, which is the typical distance at which the region heated by a point-like source is distorted by the flow. For a Keplerian flow, and a planet with small eccentricity or inclination, it is given by \citep{2017MNRAS.472.4204M,2019MNRAS.485.5035F}:
\begin{equation}
  \label{eq:1}
  \lambda = \sqrt{\frac{\chi}{(3/2)\Omega_p\gamma}}
 \end{equation}
  where  $\Omega_p$ is the planet's orbital frequency, $\chi$ is the thermal diffusivity of the gas and $\gamma$ its adiabatic index. \citet{2019MNRAS.485.5035F} obtained the expression of the thermal force onto an eccentric and inclined  planet as a function of time when two conditions are verified: 
  (i) the epicyclic and vertical excursions are small compared to the thermal lengthscale, i.e.:
  \begin{equation}
    \label{eq:2}
    ea \ll \lambda \mbox{~~~and~~~} ia \ll \lambda,
  \end{equation}
  where $e$ and $i$ are respectively the eccentricity and inclination, and $a$ is the semi-major axis of the planet; and (ii) the offset $x_p$ between the planet's corotation and its orbit (which arises from the fact that a gaseous disc is slightly sub-Keplerian) is much smaller than the planet's epicyclic and vertical excursions:
  \begin{equation}
    \label{eq:3}
    |x_p| \ll ea\mbox{~~~and~~~}|x_p| \ll ia.
  \end{equation}
When these hypothesis are satisfied, linear theory indicates that the components of the force exerted by the disc onto the planet are given by:
\begin{align}
    F_{x} & = e F_{0} \left( f_{x}^{c} \cos \Omega_pt + f_{x}^{s} \sin \Omega_pt \right) \label{eq:4}\\
    F_{y} & = e F_{0} \left( f_{y}^{c} \cos \Omega_pt + f_{y}^{s} \sin \Omega_pt \right) \label{eq:5}\\
    F_{z} & = i F_{0} \left( f_{z}^{c} \cos \Omega_pt' + f_{z}^{s} \sin \Omega_pt' \right)\label{eq:6} 
\end{align}
where $t$ is the time elapsed since a passage at periastron and $t'$ the time elapsed since a passage at an ascending node. In these expressions, we use the conventional notation for the axes: $x$ is directed along the gradient of unperturbed velocity, $y$ is directed along the unperturbed motion, and $z$ is perpendicular to the disc's midplane. In Eqs.~\eqref{eq:4}--\eqref{eq:6} the factors $f_{x,y,z}^{c,s}$ are dimensionless coefficients whose determination constitutes the main part of the linear analysis and which are \citep{2019MNRAS.485.5035F}:
\begin{align}
    f_{x}^{c} = - 0.507; & \hspace{0.5cm} f_{x}^{s} = + 1.440 \label{eq:7} \\
    f_{y}^{c} = + 0.737; & \hspace{0.5cm} f_{y}^{s} = + 0.212 \label{eq:8} \\
    f_{z}^{c} = + 1.160; & \hspace{0.5cm} f_{z}^{s} = + 0.646 \label{eq:9},
\end{align}
while $F_0$ is a normalisation factor that is the sum of two contributions:
\begin{equation}
  \label{eq:10}
  F_0 = F_0^\mathrm{cold}+F_0^\mathrm{heating},
\end{equation}
with
\begin{equation}
    F_{0}^{\text{cold}}  = - \frac{ \gamma^{ \frac{3}{2} } \left( \gamma - 1 \right) G m a L_{c} \left( \frac{3}{2} \Omega_p \right)^{ \frac{1}{2}}}{2 \pi c_{s}^{2} \chi^{\frac{3}{2}} } \label{eq:11}
\end{equation}   
and 
\begin{equation}   
    F_{0}^{\text{heating}}  = \frac{ \gamma^{ \frac{3}{2} } \left( \gamma - 1 \right) G m a L \left( \frac{3}{2} \Omega_p \right)^{ \frac{1}{2}}}{2 \pi c_{s}^{2} \chi^{\frac{3}{2}} },  \label{eq:12}
  \end{equation}
  where $G$ is the gravitational constant, $m$ is the mass of the planet, $L$ its luminosity, $c_s$ the adiabatic speed of sound and $L_c$ is given by:
  \begin{equation}
    \label{eq:13}
    L_{c} = \frac{4 \pi G m \chi \rho_{0}}{\gamma},
  \end{equation}
  where $\rho_0$ the gas density in the midplane of the disc. In what follows we will assess whether the force obtained in numerical simulations is indeed compatible with that given by Eqs.~\eqref{eq:4}--\eqref{eq:6}. We note from Eqs.~\eqref{eq:10}--\eqref{eq:12} that $L_c$ appears as the critical luminosity at which the thermal force reverses its direction, as mentioned in the Introduction. We will check this expectation with dedicated simulations.

 When the eccentricity and inclination are small, the time derivative of the orbital elements is \citep[e.g.][]{1976AmJPh..44..944B}:
\begin{align}
    \frac{de}{dt} & = \frac{1}{ m a \Omega_p} \left( F_{x} \sin \Omega_pt + 2 F_{y} \cos \Omega_pt \right) \label{eq:14}\\
    \frac{di}{dt} & = \frac{F_z}{ m a \Omega_p} \cos \Omega_pt'  \label{eq:15}\\
    \frac{d \psi }{dt} & = \frac{F_z}{ m a i \Omega_p} \sin \Omega_pt'\label{eq:16}\\
    \frac{d \varpi }{dt} & = \frac{1}{ m a e \Omega_p} \left( - F_{x} \cos \Omega_pt + 2 F_{y} \sin \Omega_pt \right).\label{eq:17}
\end{align}
Time-averaging these expressions over one orbital period and using Eqs.~\eqref{eq:4} to~\eqref{eq:13}, we obtain:
\begin{alignat}{3}
    & \frac{1}{e} \left\langle \frac{d e }{dt} \right\rangle && = \frac{\ell - 1 }{ \tau_{\text{thermal}}} \left( \frac{f_{x}^{s}}{2} + f_{y}^{c} \right) && = \frac{+1.457}{ \tau_{\text{thermal}}} \left( \ell - 1 \right) \label{eq:18}\\
    & \frac{1}{i} \left\langle \frac{d i}{dt} \right\rangle && = \frac{ \ell - 1}{\tau_{\text{thermal}}} \left( \frac{f_{z}^{c}}{2} \right) && = \frac{+0.580}{\tau_{\text{thermal}}} \left( \ell - 1 \right) \label{eq:19}\\
    & \left\langle \frac{d \psi }{dt} \right\rangle && = \frac{ \ell - 1 }{\tau_{\text{thermal}}} \left( \frac{f_{z}^{s}}{2} \right) && = \frac{+0.323}{\tau_{\text{thermal}}} \left( \ell - 1 \right) \label{eq:20}\\
    & \left\langle \frac{d \varpi }{dt} \right\rangle && = \frac{ \ell - 1 }{\tau_{\text{thermal}}} \left( - \frac{f_{x}^{c}}{2} + f_{y}^{s}\right) && = \frac{+0.466}{ \tau_{\text{thermal}}} \left( \ell - 1 \right) \label{eq:21}
\end{alignat}
where the thermal time $\tau_{\text{thermal}}$ is defined as
\begin{equation}
  \label{eq:22}
      \tau_{\text{thermal}} \equiv \frac{c_{s}^{2} \Omega_p \lambda }{2 \left( \gamma - 1 \right) G^{2} m \rho_{0} }
\end{equation}
and 
\begin{equation}
  \label{eq:23}
    \ell \equiv \frac{L}{L_{c}}.
\end{equation}
The equations~\eqref{eq:18} and~\eqref{eq:19} show that when $\ell - 1$ is negative, the eccentricity and inclination decrease exponentially. On the other hand, when $\ell - 1$ is positive, they grow exponentially, as long as the assumptions of equation~\eqref{eq:2} are verified. Note that the drift rates of $\psi$ and $\varpi$ given by equations~\eqref{eq:20} and~\eqref{eq:21} are not the net precession rates of the periapsis and line of nodes, respectively: these axis already precess on a retrograde motion under the action of the unperturbed disc's potential at a rate independent of the planet mass \citep[][and refs. therein]{2004ApJ...602..388T}. The drift rates reported here are a minute correction to this effect for low-mass planets.

\section{Methods}
\label{sec:setup}
\subsection{The code}
\label{sec:code}
We use the hydrodynamic code FARGO3D \citep{2016ApJS..223...11B} with orbital advection enabled \citep{fargo2000}. We use a spherical mesh $\left( r, \theta, \phi \right)$ where $r$ is the distance to the star, $\theta$ is the polar angle ($\theta = \pi / 2$ at the midplane of the disc) and $\phi$ is the azimuthal angle. FARGO3D numerically solves the equations of motion of a  three-dimensional, non-self-gravitating inviscid gaseous disc. These are the continuity equation:
\begin{align}
    \partial_{t} \rho + \mathbf{\nabla} \cdot \left( \rho \mathbf{v} \right) = 0,\label{eq:24}
\end{align}
the equation of conservation of momentum:
\begin{align}
    \partial_{t} \left(\rho \mathbf{v} \right) + \mathbf{\nabla} \cdot \left( \rho \mathbf{v} \otimes \mathbf{v} + p \mathbf{I} \right) = - \nabla p - \rho \nabla \Phi\label{eq:25}
\end{align}
and the equation of evolution of the density of internal energy $\varepsilon$, which reads
\begin{align}
    \partial_{t} \varepsilon + \mathbf{\nabla} \cdot \left( \varepsilon \mathbf{v} \right) = - p \mathbf{\nabla} \cdot \mathbf{v}  - \mathbf{\nabla} \cdot \mathbf{F}_{H} + S.\label{eq:26}
\end{align}
In these equations, $\rho$, $\mathbf{v}$, and $\Phi$ represent the density and velocity of the gas and the gravitational potential, respectively. The source term for energy (arising from the heat release of the planet) is denoted with $S$ and $\mathbf{I}$ represents the unit tensor.
Finally, $\mathbf{F}_{H}$ is the heat flux which is  included in the code in an additional source step (using the operator-splitting technique) corresponding to the following differential equation \citep{2021MNRAS.501...24C}:
\begin{align}
    \partial_{t} \varepsilon  = \mathbf{\nabla} \cdot \mathbf{F}_{H} \label{eq:27}
\end{align}
where $\mathbf{F}_{H}$ is given by
\begin{align}
    \mathbf{F}_{H} = - \chi \rho \mathbf{\nabla} \left( \frac{\varepsilon}{\rho} \right).\label{eq:28}
\end{align}
\begin{table*}
	\centering  
	\caption{Characteristics of our computational meshes.}
	\label{tab:mesh1}
\begin{tabular}{cccccc}
\hline  
Direction & Number of cells & Lower boundary & Upper boundary & Extent in $H$ & Extent in $\lambda$ \\
\hline  
\hline  
\multicolumn{6}{c}{\begin{tabular}[c]{@{}c@{}}Setup for eccentric planets (half disc, $91$ cells$/H$, $7.9$ cells$/\lambda$, resolution $5.5\times 10^{-4}a$) \end{tabular}} \\
\hline                            
$r$ & $816$ & $0.775$ & $1.225$ & $9$ & $103$ \\
$\phi$ & $908$ & $-0.250$ & $0.250$ & $10$ & $115$ \\
$\theta$ & $136$ & $\pi/2-0.075$ & $\pi/2$ & $1.5$ & $17$\\
\hline  
\multicolumn{6}{c}{\begin{tabular}[c]{@{}c@{}}Setup for inclined planets (entire disc,  $71$ cells$/H$, $6.2$ cells$/\lambda$, resolution $7\times 10^{-4}a$)\end{tabular}} \\
\hline                              
$r$ & $646$ & $0.825$ & $1.275$ & $9$ & $103$\\
$\phi$ & $709$ & $-0.250$ & $0.250$ & $10$ & $115$\\
$\theta$ & $214$ & $\pi/2-0.075$ & $\pi/2+0.075$ & $3$ & $34$\\
\hline  
\end{tabular}
\end{table*}

\subsection{Protoplanetary disc}
\label{sec:protoplanetary-disc}
The equation of state we use is:
\begin{align}
  \label{eq:29}
    p = \left( \gamma - 1 \right) \varepsilon
\end{align}
where $p$ is the gas pressure. In our experiments, the adiabatic index takes the value $\gamma = 7/5$, appropriate for the diatomic molecules that constitute most of the gas in protoplanetary discs. In order to allow a comparison with analytic results, we adopt a constant thermal diffusivity $\chi$. Except in adiabatic experiments, the value of $\chi$ that we use is $\chi = 4.0 \times 10^{-5}a^2\Omega_p$.

We assume different quantities of the disc to be power laws of the radial distance $r$. The first one is the aspect ratio $h \equiv H/r$, where $H$ is the vertical scale height of the disc which obeys the power law:
\begin{align}
    h = h_{p} \left( \frac{r}{a} \right)^{f}\label{eq:30}
\end{align}
where $f$ is the flaring index and $h_{p} $ is the aspect ratio at the planet's location. In all simulations presented here, we have $h_{p} = 0.05$.
Similarly, the surface density is also chosen to be a power law of $r$:
\begin{align}
    \Sigma = \Sigma_{0} \left( \frac{r}{a} \right)^{- \alpha}\label{eq:31}
\end{align}
where $\Sigma_{0}$ is the surface density at $r = a$. In all simulations presented here, we used $\Sigma_{0} = 2 \times 10^{-3}M_\ast /(\pi a^2)$ ($M_\ast$ being the mass of the star). As we need to control the distance of the planet's guiding centre to its corotation to ensure that Eq.~\eqref{eq:3} is satisfied, and subsequently relaxed to explore regimes outside the scope of the linear analysis of \citet{2019MNRAS.485.5035F}, we have varied the value of $\alpha$ according to our needs. Its value will be specified in each section.

We consider the disc to be inviscid. This assumption should not have an impact on our results.

\subsection{Planet and stellar potentials}
\label{sec:plan-stell-potent}
The gravitational potential $\Phi$ in Eq.~\eqref{eq:25} is due to the central star and the planet and is given by
\begin{align}
    \Phi = \Phi_{\ast} + \Phi_{p}\label{eq:32}
\end{align}
where
\begin{align}
    \Phi_{\ast} = - \frac{GM_{\ast}}{r}\label{eq:33}
\end{align}
and
\begin{align}
    \Phi_{p} = - \frac{Gm}{\left( r^{\prime^{2}} + \epsilon^{2} \right)^{\frac{1}{2}}} \label{eq:34}
\end{align}
are respectively the stellar and planetary potential. In Eq. (\ref{eq:34}) $r^{\prime} \equiv \lvert \mathbf{r} - \mathbf{r}_{p} \rvert$ is the distance to the planet and $\epsilon$ is a softening length to avoid computational problems arising from a divergence of the potential in the vicinity of the planet. Our simulations have been performed with $\epsilon = 0.02 H$. Note that we do not take into account an indirect term of the potential in Eq.~\eqref{eq:34}, firstly because its effect is negligible and secondly because, as we shall see in the next section, our mesh does not cover the full azimuthal extent $[-\pi,+\pi]$. Its gradient would therefore be discontinuous at the azimuthal boundaries of the computational domain, with potentially undesirable consequences.

The force exerted by the disc onto the planet is calculated after removing from each cell the azimuthal average of the density. This trick, when used on a complete disc, cures the problem of a spurious resonance shift arising from the planet and the disc orbiting within different potentials \citep{2008ApJ...678..483B}. When dealing with a wedge, it is imperatively required: a torque could be exerted on the planet even by an unperturbed disc, if the planet is not centred in the wedge.

In order to test the analytic predictions of section~\ref{sec:resumen-ecs}, we need to ensure that the planetary mass is below a critical value below which linear theory ceases to be valid (see section~\ref{sec:behav-with-plan}). We err on the side of caution and adopt a planetary mass $m=5\times 10^{-7}M_\ast$, which is way below that value. This is the planetary mass used in all the numerical experiments reported here.

All our simulations have a duration of $5$~orbits. This duration is sufficient to capture the trend of the evolution of the orbital elements when they are allowed to vary \citep{2017arXiv170401931E}, while it is also sufficient to reach a periodic regime for the disc's force when they are not. The planetary mass is introduced over one orbital period with a sinusoidal taper. When the planet is luminous, the heat is released using the procedure described by \citet{2017arXiv170401931E}.

\subsection{Mesh domain}
\label{sec:mesh-domain}
Due to the high computational cost of our study, we use a high resolution mesh centred around the planet, covering a large number of times the thermal lengthscale, but only $O(10)$ pressure lengthscales in the radial and azimuthal directions, significantly less than is customary in simulations of planet-disc interactions. Note that the thermal lengthscale is, for the values given in section~\ref{sec:protoplanetary-disc}:
\begin{align}
  \label{eq:35}
    \lambda \approx 4.4 \times 10^{-3}a \lesssim \frac{H}{11} .
\end{align}

In order to make our results comparable to the results of \citet{2019MNRAS.485.5035F}, we give the components of the force exerted by the disc onto the planet in the Cartesian frame specified in section~\ref{sec:resumen-ecs}.

 We designed two setups with different characteristics:
\begin{enumerate}
    \item \textbf{Setup for eccentric planets}. This setup allows us to study the forces in the orbital plane (i.e. $F_{x}$ and $F_{y}$, which depend on the eccentricity $e$). In this case, we simulated only one hemisphere of the disc to lower the computational cost and used reflecting boundary conditions at the midplane. This setup allowed us to study the change in $e$ and $\varpi$. In this setup, we have planets with a finite eccentricity and a null inclination.
    \item \textbf{Setup for inclined planets}. This setup allows us to study the force perpendicular to the plane ($F_{z}$, in function of $i$). In this case, we necessarily have to simulate the two hemispheres of the disc. This setup allowed us to analyse the change in $i$ and $\Omega$. With this setup, we have planets on inclined, circular orbits.
\end{enumerate}
The characteristics of the meshes of these two setups are given in table \ref{tab:mesh1}.  In appendix \ref{Convergencia}, we present a convergence study where we show that these resolutions for both setups are adequate to carry out our experiments.

\section{Cold and heating thermal forces}
\label{sec:forces}
In this section, we show the results of different series of simulations aimed at comparing  the thermal forces obtained in numerical simulations to those predicted by the linear theory [equations~\eqref{eq:4} to~\eqref{eq:6}], first when the conditions of equation~\eqref{eq:3} are verified and then when they are not.

To address this problem, we keep the planet on a fixed (eccentric or inclined) orbit to measure the forces. This avoids that the orbital elements vary over the course of the simulation, even if the variations should be tiny. Our procedure is as follows. For a given eccentricity or inclination, we perform three different runs: an adiabatic run (i.e. a run in which the planet is non-luminous\footnote{Adiabatic runs are performed necessarily with a non-luminous planet, as the heat released by the planet would otherwise accumulate in the cells where it is released, leading quickly to numerical issues. Besides, considering a luminous planet in such case would not have an interesting physical meaning.} , and $\chi=0$), a cold run (i.e. a run in which the planet  is again non-luminous, but thermal diffusion occurs: $\chi \ne 0$ and has the value given in section~\ref{sec:protoplanetary-disc}) and a hot run (i.e. a run with same thermal diffusivity as the cold run, and in which the planet has a non-vanishing luminosity). We then obtain three time series for the net, total force: $\mathbf{F}_\mathrm{adi}(t)$, $\mathbf{F}_{\chi,L=0}(t)$ and $\mathbf{F}_{\chi,L}(t)$. From these series, we infer the cold thermal force:
\begin{equation}
  \label{eq:36}
  \mathbf{F}_\mathrm{cold}(t) = \mathbf{F}_{\chi,L=0}(t)-\mathbf{F}_\mathrm{adi}(t) 
\end{equation}
and the heating force:
\begin{equation}
  \label{eq:37}
  \mathbf{F}_\mathrm{heat}(t) = \mathbf{F}_{\chi,L}(t)-\mathbf{F}_{\chi,L=0}(t) .
\end{equation}
Naturally, from this decomposition of the force, it is an evidence to say that the total force in the most general case (i.e. with thermal diffusion and a luminous planet), $\mathbf{F}_{\chi,L}(t)$, is simply, by construction:
\begin{equation}
  \label{eq:38}
  \mathbf{F}_{\chi,L}(t) = \mathbf{F}_\mathrm{adi}(t)+\mathbf{F}_\mathrm{cold}(t)+\mathbf{F}_\mathrm{heat}(t).
\end{equation}

\subsection{No corotation offset}
\label{sec:corot-offs-x_p}
In the first set of simulations, we embed the planet in a strictly Keplerian disc so that $x_p=0$ (the guiding centre of the planet falls exactly on corotation). This implies that the condition of equation~\eqref{eq:3} is verified, and that there is no constant term for the thermal force, which depends on the corotation offset and vanishes when the offset does \citep{2017MNRAS.472.4204M,2021MNRAS.501...24C}. The corotation offset has the expression \citep[e.g.][]{2017MNRAS.472.4204M}:
\begin{align}
  \label{eq:41}
    x_{p} = \eta h^{2} a 
\end{align}
where $\eta$ is given by:
\begin{align}
  \label{eq:42}
    \eta = \frac{\alpha - f + 2}{3}.
\end{align}
We adopt a combination of $\alpha$ and $f$ that yields a null $\eta$. We take $\alpha = -1.5$ and $f = 0.5$.
\begin{figure*}
  \centering
  \includegraphics[width=.8\textwidth]{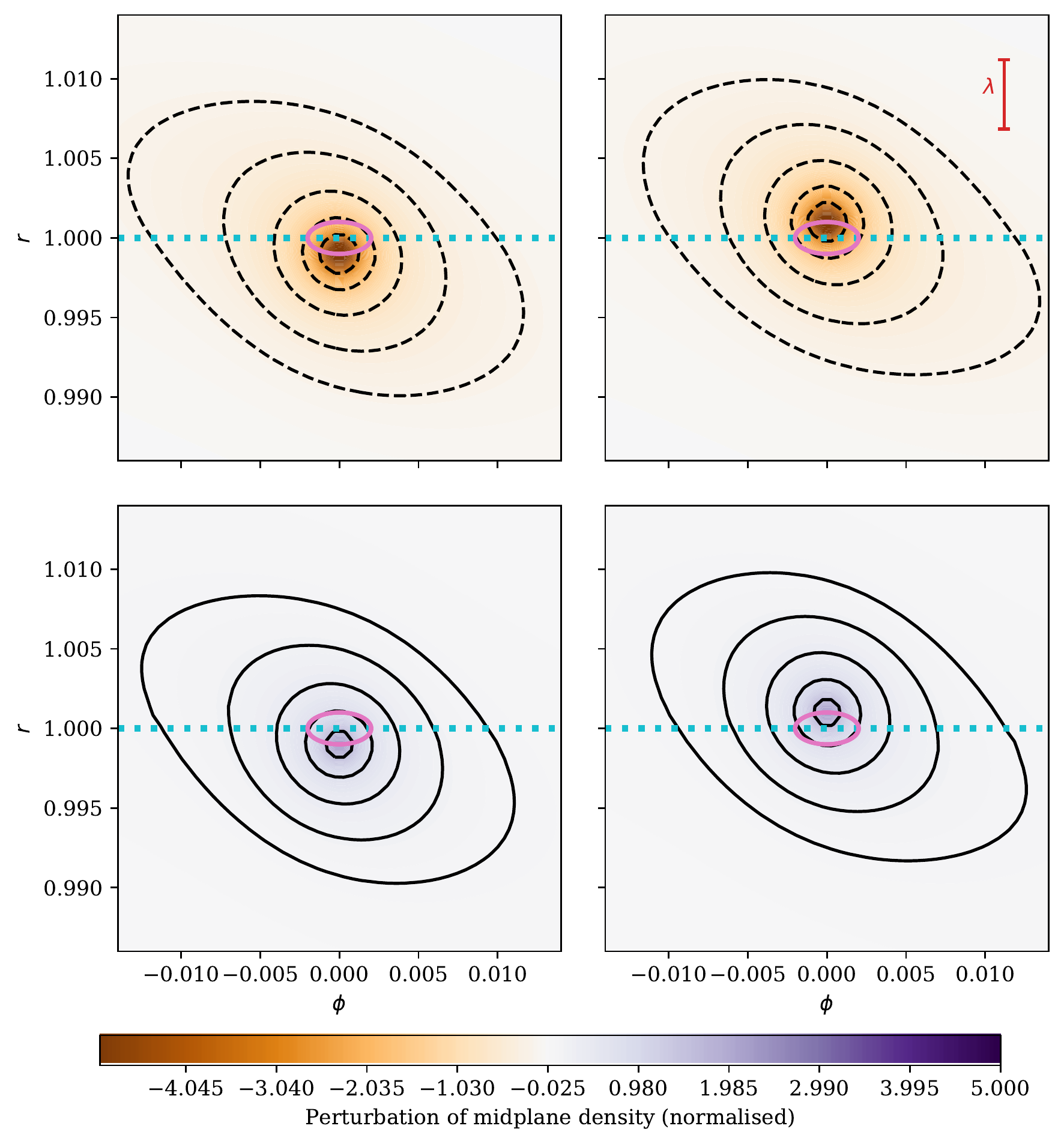}
  \caption{Perturbation of midplane density arising from the release of heat (top row) or from the occurrence of thermal diffusion (bottom row). The top row maps are obtained by subtracting the density of a cold run from that obtained in a hot run. The maps of the bottom row are obtained by subtracting the density in an adiabatic run from the density obtained in a cold run. The left (right) column shows the perturbation when the planet is at periastron (apoastron). The perturbation of the top row gives rise to the heating force, while the perturbation of the bottom row corresponds to the cold force. The horizontal dotted line shows the corotation and the ellipse shows the epicyclic trajectory of the planet, which is described anti-clockwise. Its centre falls on corotation as these runs have no corotation offset. The hot runs were performed with a planet luminosity $L=2L_c$. The density is normalised to $\gamma^{3/2}(\gamma-1)L_c\Omega_p/\chi c_s^3$. The isocontours have values $\pm 2^n/10$, with $n\in[1,5]$. The vertical segment in the top right map shows the thermal lengthscale. An animation of this figure accompanies the electronic version of this article. \label{map}}
\end{figure*}
We show in Fig.~\ref{map} the aspect of the perturbations arising from the release of heat by the planet, leading to the heating force, and that simply arising from thermal diffusion (the planet being non-luminous), leading to the cold thermal force. This perturbation resembles the two-lobe pattern due to a planet on circular orbit \citep{2017MNRAS.472.4204M,2021MNRAS.501...24C}, but we see a minute change in the shape and extent of the isocontours as a function of the orbital phase. This variation is what induces the time-varying force that we characterise in this work. We see that the introduction of thermal diffusion leads to a larger density in the vicinity of the planet with respect to the adiabatic case (the perturbation of density in the bottom plots is positive), while the release of heat in the planet's vicinity tends to lower the density.
\begin{figure}
 \includegraphics[width=\columnwidth]{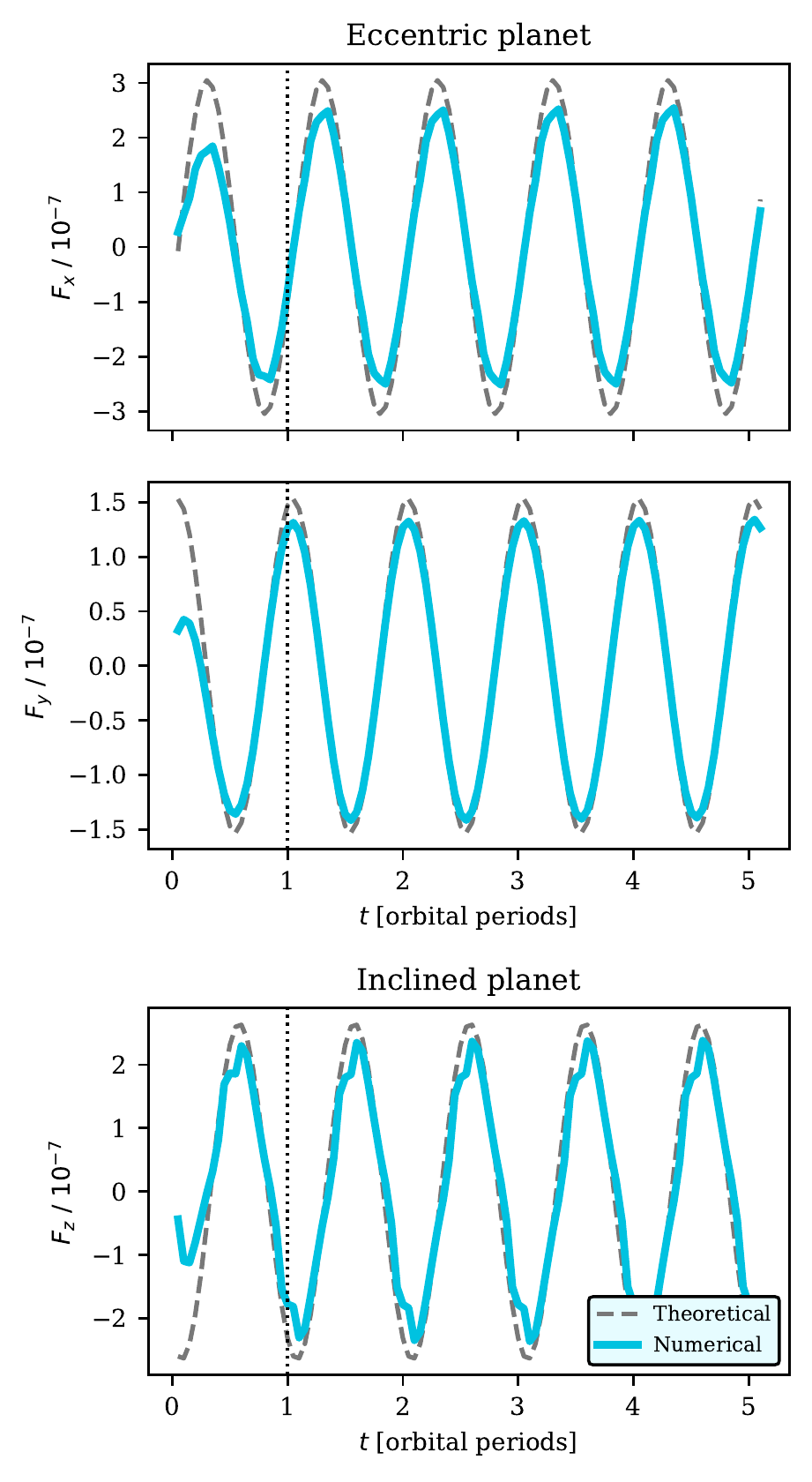}
  \caption{Components $F_{x}$ (top plot), $F_{y}$ (middle plot) of the heating force on an eccentric planet and $F_{z}$ (bottom plot) for an inclined planet. In each plot the solid blue line represents the force obtained from numerical simulations, and the grey dashed line represents the theoretical prediction of Eqs.~\eqref{eq:4} to~\eqref{eq:6}. In these runs the planet's luminosity is $L=2L_c$, and we have either $e=7.5\times 10^{-4}$ (first two plots) or $i=7.5\times 10^{-4}$ (last plot).}
 \label{fig:Ejemplo}
\end{figure}
We show examples of the heating force obtained in numerical simulations in Fig.~\ref{fig:Ejemplo}.  The top and middle plots show the  time behaviour of the horizontal components of the heating force on an eccentric planet and the bottom plot shows the vertical component of the heating force on an inclined planet. We see that the curve obtained in numerical simulations agrees reasonably well, for each component, with the theoretical prediction, both in amplitude and phase. We also see that there is a transient regime after the simulation start, which lasts approximately one orbit, after which a periodic regime is reached.

\begin{figure}
  \includegraphics[width=\columnwidth]{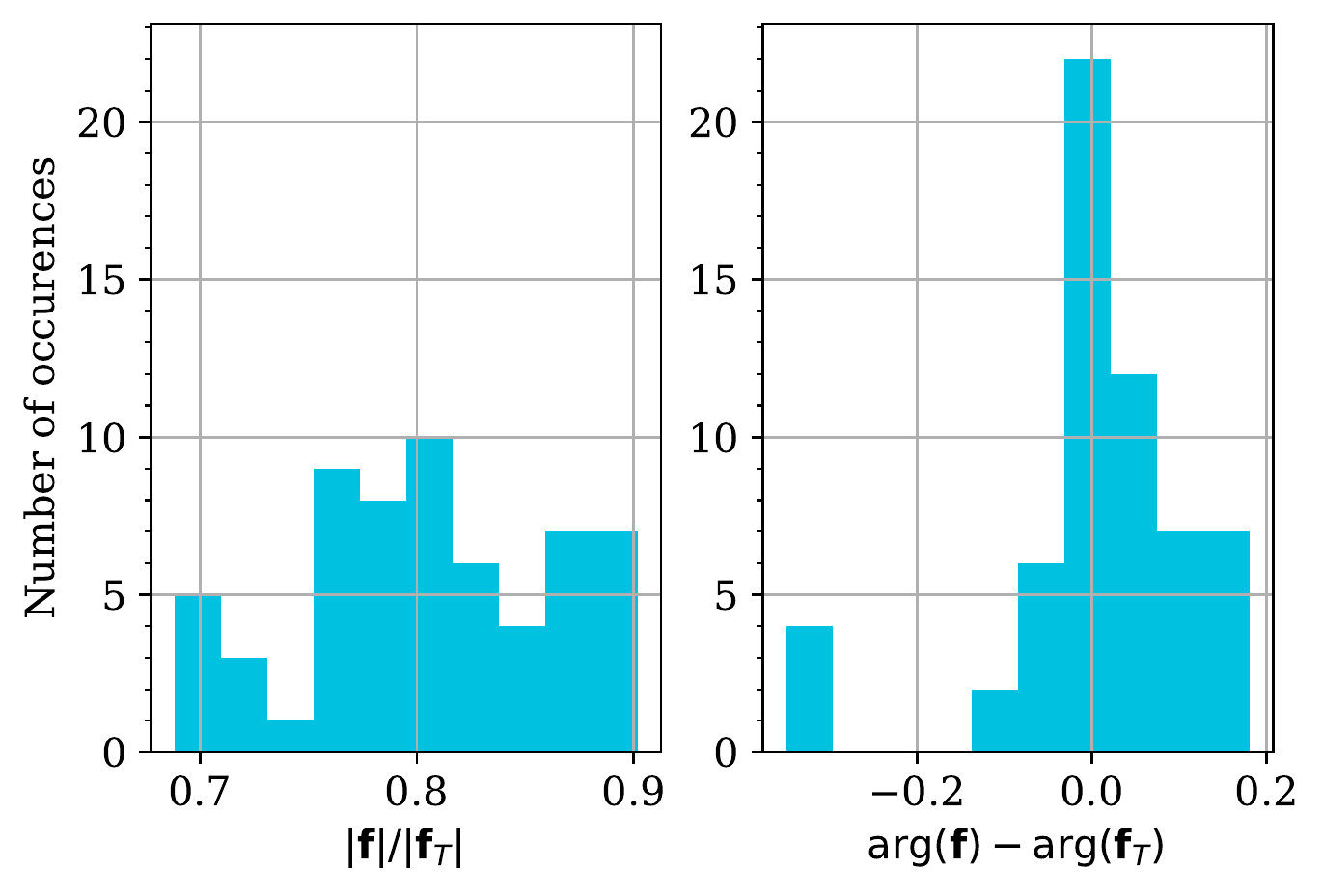}
  \caption{Histogram of the ratio of the amplitude of the oscillations of the force obtained in our numerical experiments to that expected from linear theory (left) and histogram of the phase difference (in radians) of the force oscillation between numerical experiments and predictions from linear theory (right). These histograms contain the results for the three components of the force. The isolated peak at the left of the right plot corresponds to the runs of a planet with very low inclination (see text for details).}
  \label{fig:histo}
\end{figure}

A more concise manner of comparing the numerical curves to the theoretical expectations consists in obtaining the dimensionless cosine and sine coefficients corresponding to the curves, normalised to $eF_0^\mathrm{cold/heating}$ or $iF_0^\mathrm{cold/heating}$ (see equations \ref{eq:4} to~\ref{eq:6} and equations~\ref{eq:10} to~\ref{eq:12}). Let us take the example of the $z$-component of the heating force exerted on an inclined planet. We have the time series $[F_z(j\Delta t)]_{j\in [0,100]}$, where $\Delta t= \pi/(10\Omega_p)$ is the time interval between two samplings of the force, corresponding to $1/20^{th}$ of an orbital period in all our runs. The cosine and sine coefficients are then evaluated as:
\begin{align}
  f^{c,\mathrm{num}}_z & = \frac{2}{j_1-j_0+1} \sum_{j=j_0}^{j_1} \frac{F_z(j\Delta t)}{iF_0^\mathrm{heating}} \cos \left( \Omega_p j\Delta t \right)\label{eq:39} \\
      f^{s,\mathrm{num}}_z & = \frac{2}{j_1-j_0+1} \sum_{j=j_0}^{j_1} \frac{F_z(j\Delta t)}{iF_0^\mathrm{heating}} \sin \left( \Omega_p j\Delta t \right).\label{eq:40} 
\end{align}
The starting and ending points $j_0$ and $j_1$ of the sums are chosen (i) to avoid the transient regime of the first orbit, (ii) to cover an integer number of orbital periods and (iii) to cover as large as possible a time interval to increase accuracy. We therefore choose $j_0=40$ and $j_1=99$ in order to cover the last three orbital periods.
\begin{figure*}
    \includegraphics[width=\textwidth]{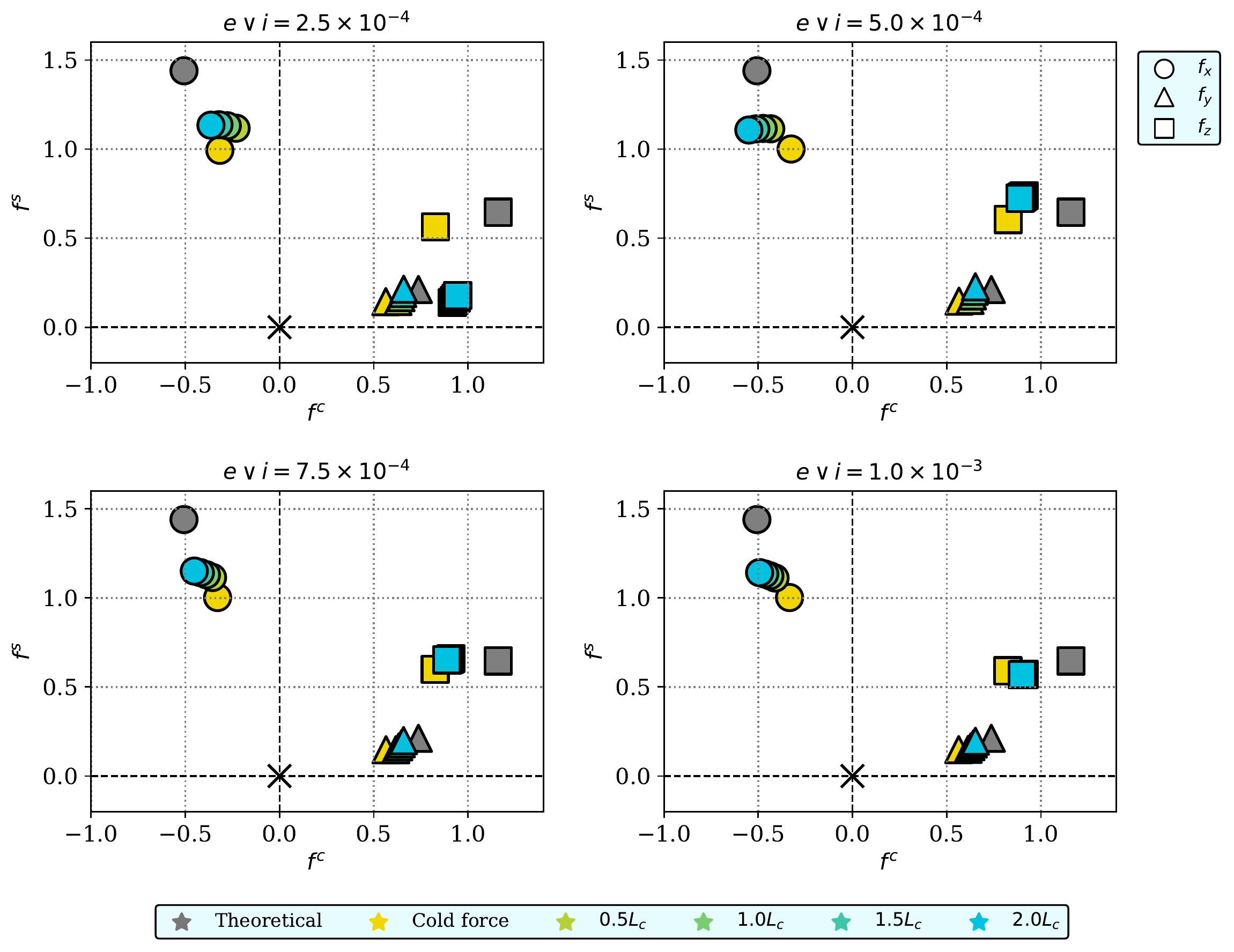}
    \caption{Representation of the force coefficients in the complex plane: circles represent the coefficients $f_{x}$, triangles the coefficients $f_{y}$, and squares the coefficients $f_{z}$. The yellow markers show the results of experiments for the cold thermal force; the others show the results for the hot thermal force using different luminosities. The grey markers represent the theoretical predictions corresponding to equations~\eqref{eq:7} to~\eqref{eq:9}.}
    \label{fig:Nube}
\end{figure*}
It is then convenient to represent the coefficients of a given force component by a point in the complex plane, with coordinates $(f^c, f^s)$, in order to compare its position to that expected from linear theory. We present our results in Fig.~\ref{fig:Nube}, which shows the points representative of the three different components of the force obtained for different values of the planet's luminosity (for the heating force) or from the cold force, and for different values of the eccentricity or inclination. Given that we have considered four values of the eccentricity or inclination, and five different cases (cold case and four different luminosities), we obtain $60$ points in total. For each of those, we can evaluate the ratio of the module of the complex number $\mathbf{f}=f^c+jf^S$ (where $j^2=-1$) to that given by linear theory $\mathbf{f}_T=f^c_T+jf^S_T$, as well as the difference of the arguments of these two numbers. Fig.~\ref{fig:histo} shows the histogram of these two quantities.
We see on this figure that no calculation gives a match of the amplitude better than $10$~\% (the highest value of $|\mathbf{f}|/|\mathbf{f}_T|$ is $0.9$) while no calculation yields an amplitude smaller than $70$~\% of that predicted. The mode of the distribution is at $80$~\%: the amplitude of the oscillations of the force components in our set of numerical experiments is in general $20$~\% smaller than expected. The agreement on the phase of the oscillations (between numerics and theory) is better, as it is at most $\sim 0.18$~rad ($\lesssim 0.03$~turn). There is one exception: for the lowest value of inclination, the vertical heating force is significantly at odds with theoretical expectations (see the square symbols in the top-left plot of Fig.~\ref{fig:Nube}). That case, however, is extreme: the vertical excursion corresponding to this low inclination is only $\sim 35$~\% of the cell size, and is probably too low to get a correct response from the flow. The minimal excursion seems to be of the order of half a cell: in that same plot, the points representative of the horizontal components of the force do not display larger offsets than at larger eccentricity, while they have a radial excursion $ea$ of exactly half a cell (see Tab.~\ref{tab:mesh1}).

\subsection{Finite corotation offsets}
\label{sec:large-corot-offs}
\begin{figure*}
    \includegraphics[width=.85\textwidth]{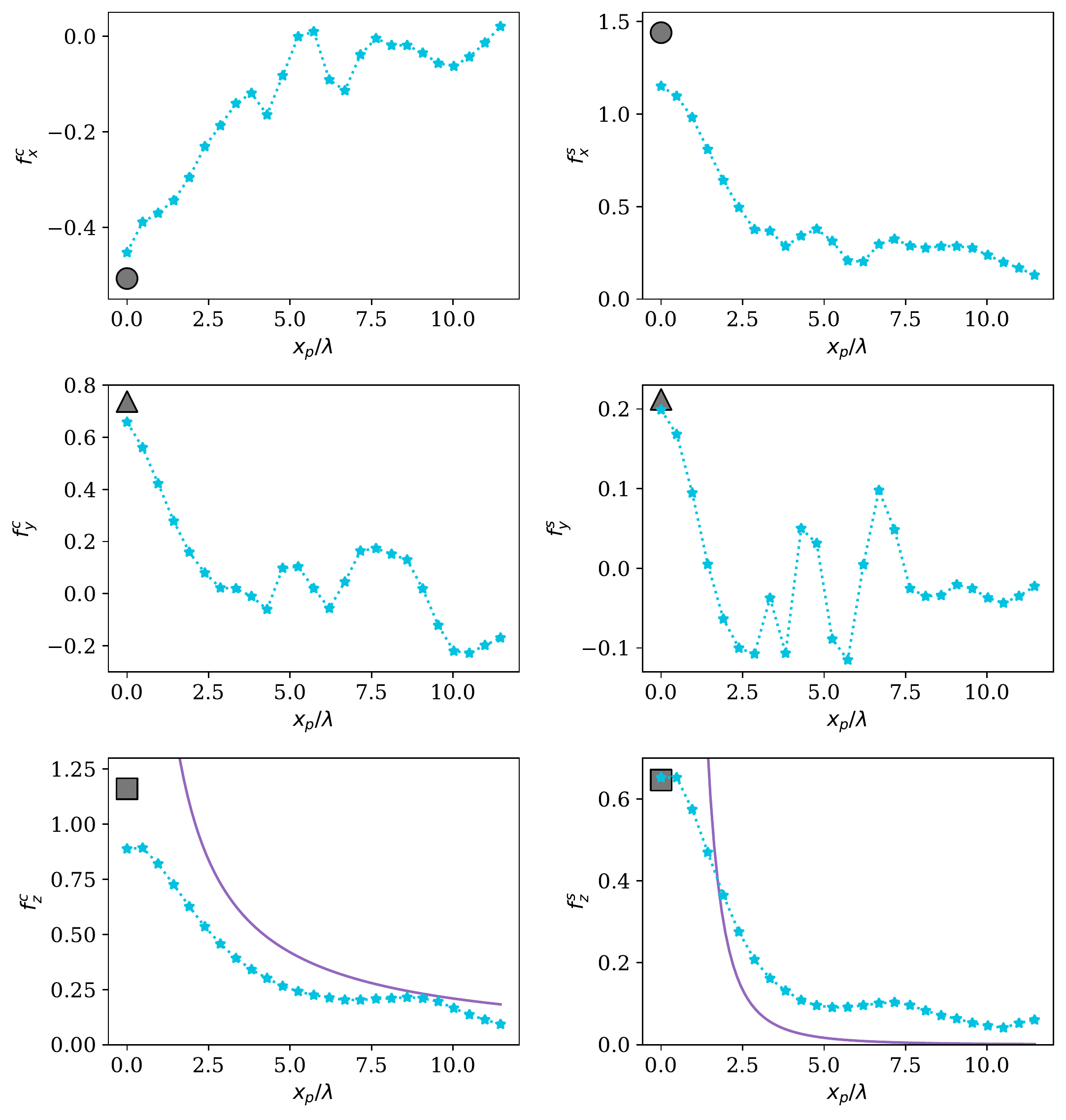}
    \caption{This figure shows the cosine and sine coefficients of the force as a function of the distance to corotation, normalised to the thermal lengthscale. These data were obtained for a planet's luminosity $L=2L_c$. The grey symbols correspond to the theoretical values for $|x_p| \ll \lambda$, given by Eqs.~\eqref{eq:7} to~\eqref{eq:9}. The purple line corresponds to the value of the cosine and sine coefficients given by Eq.~(\ref{eq:47}) and~(\ref{eq:53}), respectively.}
    \label{fig:Coeficientes-NoTanLineal}
\end{figure*}

\begin{figure}
  \begin{center}
    \includegraphics[width=.9\columnwidth]{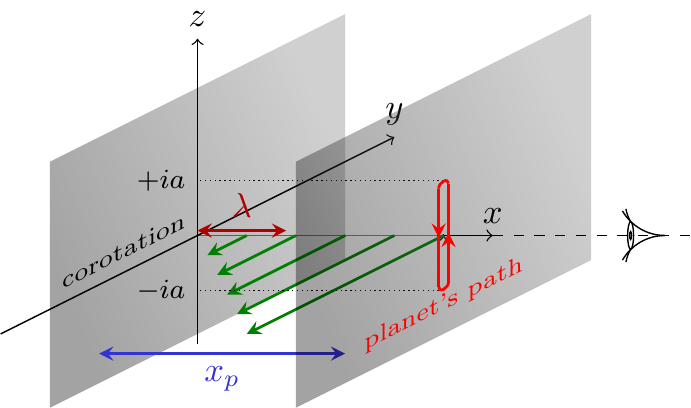}
  \end{center}
  \caption{Schematic view of the path of an inclined planet seen in the corotating frame. This path reduces to the segment $x=x_p$, $y=0$ and $|z| \le ia$. The distance $x_p$ of the planet's path to the corotation sheet is significantly larger than the thermal lengthscale $\lambda$. The green arrows depict the Keplerian sheared flow.\label{fig:scheme}}
\end{figure}

\begin{figure}
  \includegraphics[width=.5\columnwidth]{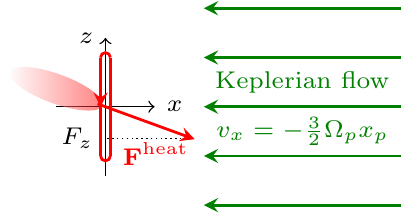} \includegraphics[width=.5\columnwidth]{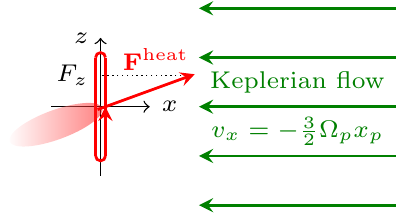}
  \caption{Depiction of the heated region downwind of the planet as seen by the observer of Fig.~\ref{fig:scheme}, when the planet moves downwards (left) and upwards (right), and of the resulting heating force (red arrow). The green arrows show the Keplerian flow in the plane $x=x_p$.}
  \label{fig:hf12}
\end{figure}

We also have explored the regime of finite corotation offsets up to large values, with runs additional to those mentioned above. Writing $x_{p} = \xi \lambda$, we have, using Eq.~\eqref{eq:41} and~\eqref{eq:42}:
\begin{equation}
\alpha - f =  \frac{3 \xi\lambda}{h^{2} a} -2 
\end{equation}
We can vary either $\alpha$ or $f$ to explore the force dependence on $\xi$. Since we wish to explore this dependence up to large values of $\xi$ (namely we go up to $\xi \gtrsim 11$), we have to take large values of $\alpha$ or $f$. The latter can lead to large sound speeds near one edge of the mesh, inducing severe restrictions on the time step. For this reason we prefer to vary $\alpha$, and keep $f$ fixed at $0.5$.  While the large values of $\alpha$ needed to reach our most extreme values of $\xi$ are not realistic, it is still of interest to study a regime where the corotation offset is significantly larger than the thermal lengthscale $\lambda$, as the conditions for such regime may be met in the earlier stages of the disc, when the latter is very opaque (which entails a small $\lambda$) and hot (so that $x_p$ is large). Here, instead of decreasing $\lambda$, which would be impractical from a resolution standpoint, we rather increase $x_p$ by imposing large, not necessarily realistic gradients of surface density.
For each setup, we perform 25 experiments where we take $x_{p} / \lambda$ from $0$ to $11.456$ (which corresponds to a value of $x_{p} = H$ for our parameters). This translates into varying $\alpha$ from $-1.5$ to $58.5$. All these experiments were carried out with an eccentricity or inclination equal to $7.5 \times 10^{-4}$. The results of these experiments are shown in Fig. \ref{fig:Coeficientes-NoTanLineal}.
We see that when $x_p > \lambda$, the coefficients are smaller than their value at $x_p=0$, and they decay towards small values at larger distance from corotation. In order to get some insight into this behaviour, we work out the thermal response to an inclined planet orbiting at a distance larger than $\lambda$ from its corotation. We depict this situation in Fig.~\ref{fig:scheme}. In this case, the response time of the force is shorter than the shear timescale, so that the force can be given by a local calculation that neglects the shear \citep{2002A&A...388..615P,2017MNRAS.465.3175M}, i.e. a calculation of dynamical friction. The heated region downstream of the planet has an inclination that depends on the vertical velocity of the planet, as shown in Fig.~\ref{fig:hf12}.
The heating force is therefore \citep{2017MNRAS.465.3175M,2019MNRAS.483.4383V}:
\begin{equation}
  \label{eq:43}
  \mathbf{F}^\mathrm{dyn}=\frac{\gamma(\gamma-1)GmL}{2\chi c_s^2} \frac{\mathbf{V}}{V}
\end{equation}
where $\mathbf{V}$ is the planet's velocity with respect to the gas, which is:
\begin{equation}
  \label{eq:44}
  \mathbf{V} = \left[+\frac{3}{2}\Omega_px_p,0,ia\Omega_p\cos(\Omega_pt)\right]^T.
\end{equation}
We assume that the tilt angle of the hot plume (and heating force) with respect to the horizontal plane is small, which entails that $ia\Omega_p\ll (3/2)\Omega_px_p$, or equivalently that $i \ll (3/2) \eta h^2$. This condition is fulfilled in our setup. This implies that the vertical component of the force :
\begin{equation}
  \label{eq:45}
  F^\mathrm{dyn}_z=\frac{\gamma(\gamma-1)GmL}{3\chi c_s^2}\frac{ia\cos(\Omega_pt)}{x_p}.
\end{equation}
Using Eqs.~\eqref{eq:1} and~\eqref{eq:12}, we can rewrite the above equation as:
\begin{equation}
  \label{eq:46}
  F^\mathrm{dyn}_z=iF_0^\mathrm{heating}\frac{2\pi}{3}\frac{\lambda}{x_p}\cos(\Omega_pt). 
\end{equation}
By comparing this result to Eq.~\eqref{eq:6}, we read directly the cosine coefficient of the vertical force for a large corotation offset:
\begin{equation}
  \label{eq:47}
  f_z^C = \frac{2\pi}{3}\frac{\lambda}{x_p}.
\end{equation}
This value is shown as the solid purple line in the left bottom plot of Fig.~\ref{fig:Coeficientes-NoTanLineal}. We see that it follows the trend displayed by the coefficient obtained from numerical simulations, although the latter is systematically smaller. This smaller value in numerical simulations is attributable to the resolution. The size of the hot trail following the planet is \citep{2017MNRAS.465.3175M}:
\begin{equation}
  \label{eq:48}
  \lambda^\mathrm{dyn}=\frac{\chi}{\gamma V},
\end{equation}
where we use the superscript dyn to avoid confusion with the value given by Eq.~\eqref{eq:1}, which applies to thermal disturbances sheared by the flow. For $x_p/\lambda=5$, we work out: $\lambda^\mathrm{dyn}\approx 1.5\times 10^{-3}a \approx 2.1$~cells only. While our resolution is sufficient to capture the force at $x_p/\lambda \sim 0$, it is not enough to provide an accurate  value of the latter at large corotation offsets, and the situation worsens as the corotation offset increases, by virtue of Eq.~\eqref{eq:48}.

Eq.~\eqref{eq:46} seems to suggest that the sine coefficient of the force is zero. This happens because the estimate above assumes that the force reacts instantaneously to a change in velocity. We can improve our estimate by introducing the response time of the force \citep{2017MNRAS.465.3175M}:
\begin{equation}
  \label{eq:49}
  \tau = \frac{\chi}{\gamma^2V^2}.
\end{equation}
Eq.~\eqref{eq:43} then becomes:
\begin{equation}
  \label{eq:50}
    \mathbf{F}^\mathrm{dyn}(t)=\frac{\gamma(\gamma-1)GmL}{2\chi c_s^2} \left.\frac{\mathbf{V}}{V}\right|_{t-\tau},
  \end{equation}
  so that Eq.~\eqref{eq:46} becomes:
  \begin{equation}
    \label{eq:51}
  F^\mathrm{dyn}_z=iF_0^\mathrm{heating}\frac{2\pi}{3}\frac{\lambda}{x_p}\cos(\Omega_pt-\Omega_p\tau).     
\end{equation}
Assuming that $\tau \ll \Omega_p^{-1}$, we write $\cos(\Omega_pt-\Omega_p\tau)=\cos(\Omega_pt)+\Omega_p\tau\sin(\Omega_pt)$. Using $V=(3/2)\Omega_px_p$ and using Eq.~\eqref{eq:1}, we obtain:
\begin{equation}
  \label{eq:52}
  F^\mathrm{dyn}_z=iF_0^\mathrm{heating}\frac{2\pi}{3}\left[\frac{\lambda}{x_p}\cos(\Omega_pt)+\left(\frac{\lambda}{x_p}\right)^3\sin(\Omega_pt)\right].
\end{equation}
The sine coefficient of the force is therefore:
\begin{equation}
  \label{eq:53}
  f_z^S=\frac{2\pi}{3}\left(\frac{\lambda}{x_p}\right)^3.
\end{equation}
It decays faster than its cosine counterpart, as the response time of the force decreases at larger velocity (i.e. at larger corotation offset). The dependence given by Eq.~\eqref{eq:53} is plotted as a purple solid line on the right bottom plot of Fig.~\ref{fig:Coeficientes-NoTanLineal}. The decay observed in numerical experiments is not as sharp as that given by Eq.~\eqref{eq:53}. The very limited resolution at large $x_p$ prevents the response time to decay to values smaller than $\Delta x/V$. Yet our simple modelling gives a qualitative explanation of why the sine coefficient decreases faster than the cosine one.

\subsection{Cancellation of the thermal force}
\label{sec:canc-therm-force}
\begin{figure}
 \includegraphics[width=\columnwidth]{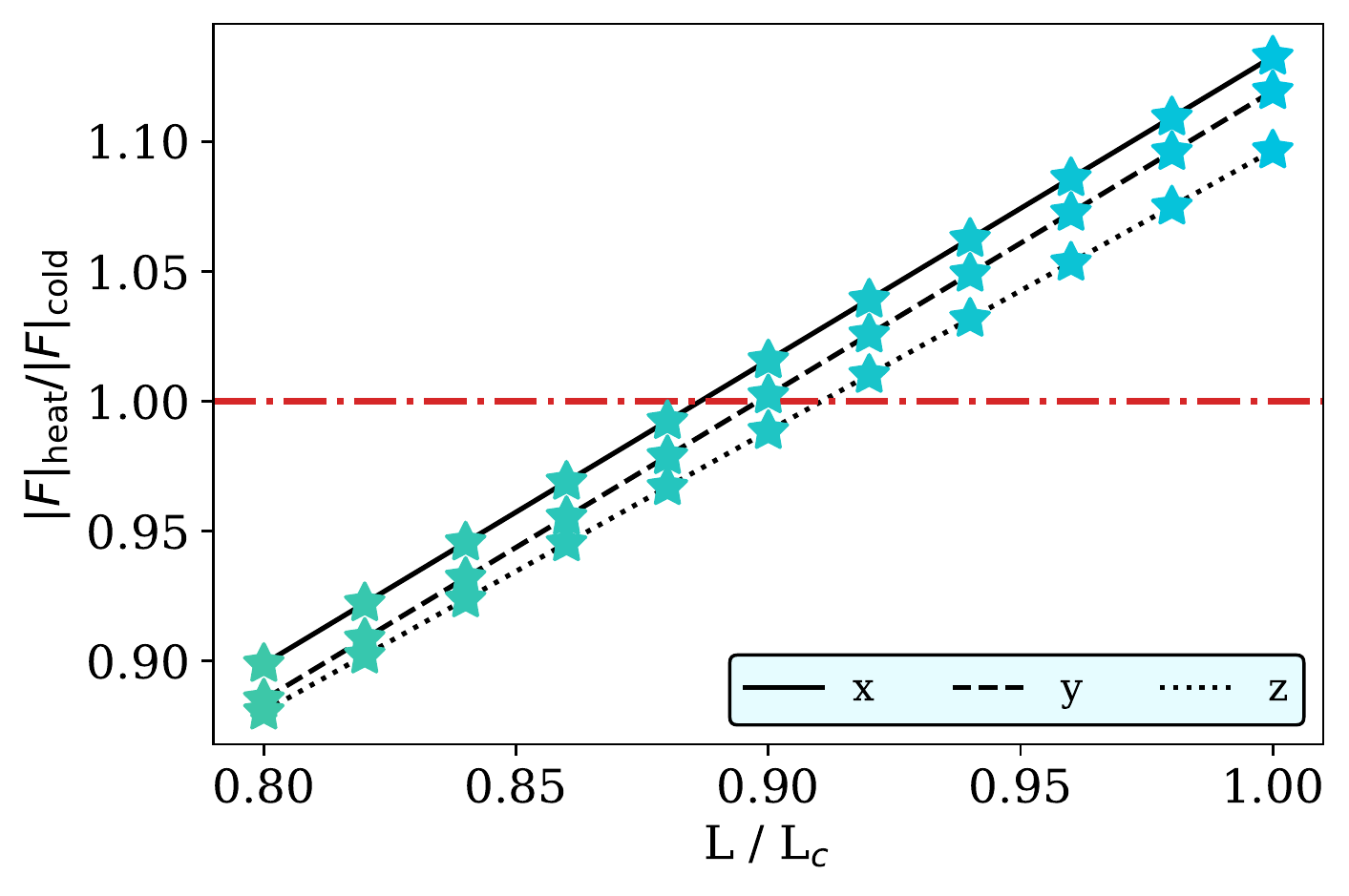}
 \caption{The ratio between the amplitude of the heating force to that of the cold force, for each of the three components of the force. The amplitude is determined as $|F| = \left[(F^c)^2+(F^s)^2\right]^{1/2}$. The red line shows where the ratio equals 1. The component of net thermal force cancels out when the ratio is one. We see that the three components do  not cancel exactly for the same luminosity, but for three very close values, which are $ 0.886 L_{c}$, $L = 0.898 L_{c}$,  $0.911 L_{c}$ for $F_x$, $F_y$ and $F_z$, respectively.}
 \label{fig:std-Adiabatico}
\end{figure}

In this section, we look for the luminosity value required to cancel the net heating force. From Eqs.~\eqref{eq:11} and~\eqref{eq:12}, we expect the cancellation to occur when $L=L_c$. In that case, the net force onto the planet, from Eq.~\eqref{eq:38}, reduces to the force of the adiabatic case (on a non-luminous planet). From the set of experiments we have carried out so far, we see that the normalised amplitude for the cold force is systematically slightly smaller than that of the heating force. We therefore expect the cancellation of the net thermal force to occur for $L\lesssim L_c$. Therefore, we carry out, both with the setup of eccentric planets and the setup of inclined planets, 1) eleven simulations varying the luminosity of the planet between $0.8L_c$ and $1.0L_{c}$, 2) a simulation with a non-luminous planet, 3) a simulation with a non-luminous planet in an adiabatic disc. By subtracting the force obtained in case 2) from case 3), we obtain the cold force. By subtracting the forces obtained in case 1) from case 2), we obtain the values of the heating force for the different luminosities considered. We plot the ratio of the heating force's amplitude to that of the cold force as a function of the luminosity in Fig.~\ref{fig:std-Adiabatico}, for the three components of the force. 
We see that the cancellation of the thermal force is attained for a planet luminosity $\sim 10$~\% smaller than the critical luminosity predicted by linear theory.
Numerical simulations of thermal forces on a planet on circular orbit have reached a similar conclusion: the luminosity required to cancel out the net force is slightly below $L_c$ \citep{2021MNRAS.501...24C}.  This may be due to the resolution used in our simulations as well as to the use of a softening length of the potential (which represents $\sim 1.8$~cells in runs with eccentric planets, and $\sim 1.4$~cells in runs with inclined planets: these values are not very small compared to the thermal lengthscale, as can be seen in Tab.~\ref{tab:mesh1}). The cold force arises from the excess of cold gas that floods the potential well of the planet, with respect to the adiabatic situation in which there is much less gas near the planet. The analytic estimate of the cold force assumes that the potential is exactly that of a point-like mass. This is not true when using a softening length, and there is therefore less mass near the planet, which can reduce the value of the cold force. Yet the impact of the finite resolution and of the softening length are relatively minor, so that the luminosity required to cancel the net thermal force is in good agreement with the prediction of  \citet{2019MNRAS.485.5035F}.

\section{Variation of the orbital elements}
\label{sec:orbitalelements}
In this second part of the article, we relax the assumption of section~\ref{sec:forces} that the planet is held on a fixed eccentric or inclined orbit, and we let it evolve freely under the action of the star's and disc's forces in order to evaluate the variation rate of its orbital elements. The setups that we use to carry out these simulations are identical to the ones we used in the previous section, except that we allow the planet to sense the disc's force, allowing its orbital elements to change.

The eccentricity and inclination vary independently, for the small values that we consider here. We could study their variations with one run with an eccentric and inclined planet. This would imply that the eccentricity's variation rate is studied with the coarser resolution of the full disc setup (see section~\ref{sec:mesh-domain}). For this reason, we prefer to study the variations of eccentricity and inclination in different runs, so as to take advantage of the higher resolution of the half-disc setup.
As in section~\ref{sec:corot-offs-x_p}, we use a null corotation offset.

\begin{figure*}
 \includegraphics[width=.85\textwidth]{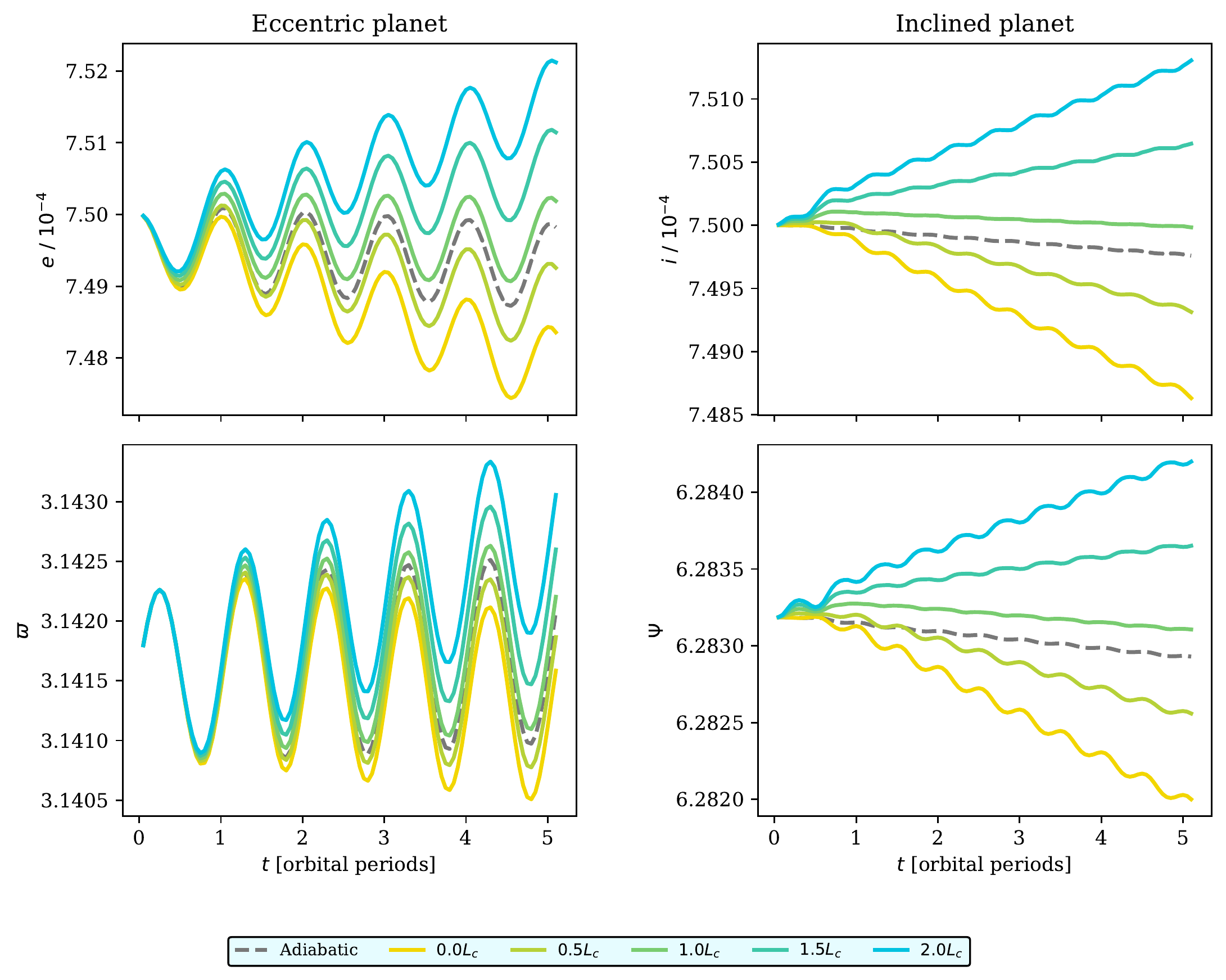}
 \caption{Evolution of the eccentricity $e$, inclination $i$, longitude of ascending node $\Psi$ and longitude of periapsis $\varpi$ of luminous planets. The colour of the line indicates the value of the luminosity. The grey dashed lines show the evolution of the orbital elements of a (non-luminous) planet embedded in an adiabatic disc.}
 \label{fig:Cambio_Elementos}
\end{figure*}

\begin{figure*}
 \includegraphics[width=.9\textwidth]{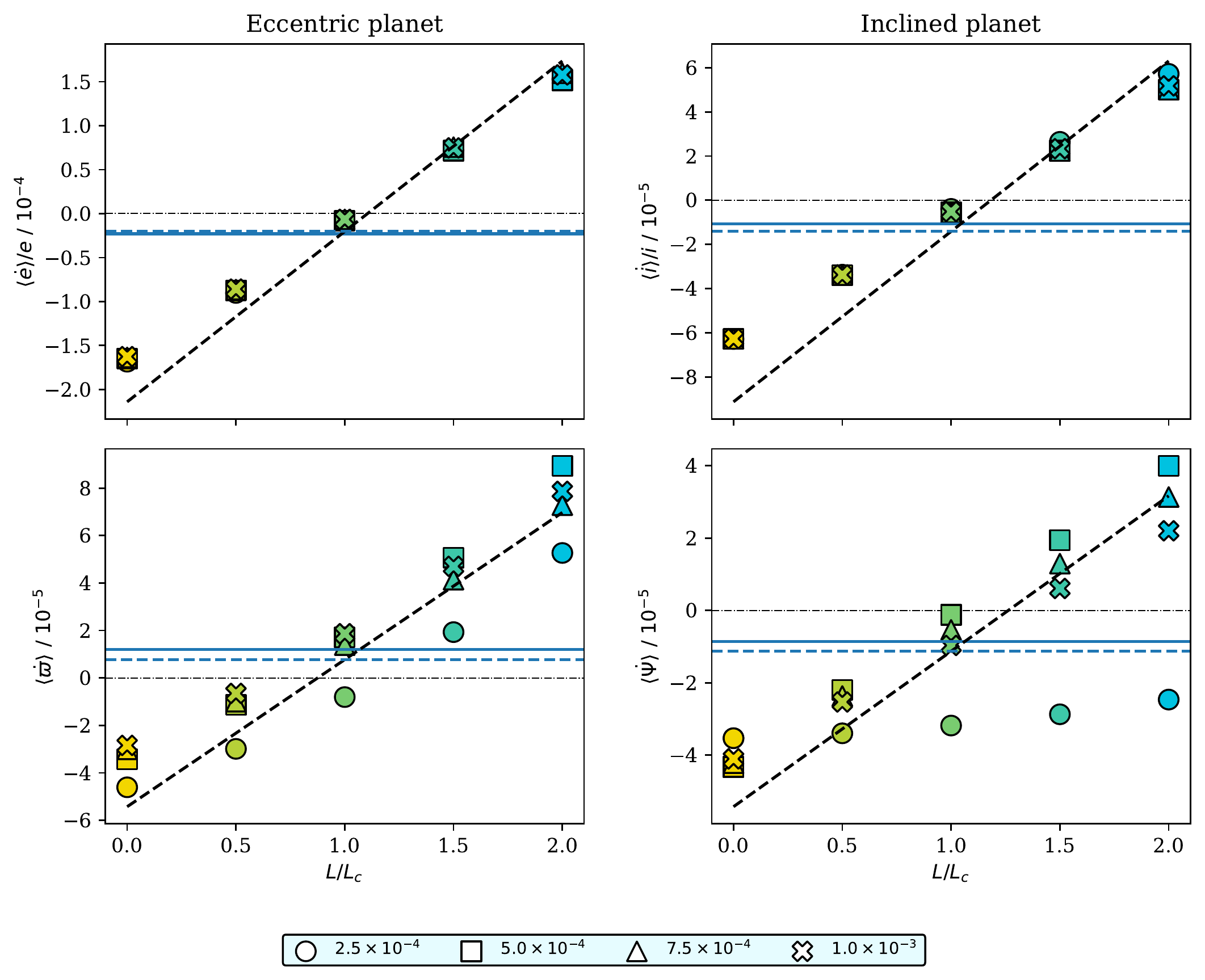}
 \caption{Mean evolution rates of eccentricity $e$, inclination $i$, longitude of periapsis $\varpi$ and longitude of ascending node $\Psi$ as a function the planet's luminosity. The shape of the marker indicates the initial value of the eccentricity or inclination, depending on the simulation. The horizontal blue lines show the evolution rate of the orbital elements in an adiabatic disc (for a non-luminous planet), as a reference. The solid lines correspond to the rates measured in the numerical experiments, whereas the dashed lines correspond to the rates given by \citep{2004ApJ...602..388T}. The tilted dashed lines represent the dependencies from Eqs.~\eqref{eq:54} to~\eqref{eq:57}.}
 \label{fig:Cambio_Elementos-Coeficientes}
\end{figure*}

Fig.~\ref{fig:Cambio_Elementos} shows the variation of the orbital elements for $e \vee i = 7.5 \times 10^{-4}$.
Unlike our study of section~\ref{sec:forces}, we do not separate here the action of thermal forces from that resulting from the interaction with the disc at Lindblad's and corotation resonances. As such, the change rate of the orbital elements in our runs also incorporates the variations arising from wave-launching. From linear theory, they should be:
\begin{align}
    \frac{1}{e} \left\langle \frac{d e }{dt} \right\rangle & = \frac{1.457}{ \tau_{\text{thermal}} } \left( \ell - 1 \right) - \frac{0.780}{ \tau_{\text{wave}} }\label{eq:54}\\
  \frac{1}{i} \left\langle \frac{d i}{dt} \right\rangle & = \frac{0.580}{ \tau_{\text{thermal}} } \left( \ell - 1 \right)  - \frac{0.544}{ \tau_{\text{wave}} }\label{eq:55}\\
 \left\langle \frac{d \Omega }{dt} \right\rangle & = \frac{0.323}{ \tau_{\text{thermal}} } \left( \ell - 1 \right)  - \frac{0.435}{ \tau_{\text{wave}} } \label{eq:56}\\
 \left\langle \frac{d \varpi }{dt} \right\rangle & = \frac{0.466}{ \tau_{\text{thermal}} } \left( \ell - 1 \right)  + \frac{0.297}{ \tau_{\text{wave}} }\label{eq:57}
\end{align}
where $\ell$ is given by Eq.~\eqref{eq:23} and $\tau_{\text{wave}}$ is given by \citep{2004ApJ...602..388T}:
\begin{align}
  \tau_{\text{wave}} &= \left(\frac{m}{M_\ast}\right)^{-1}\left(\frac{\Sigma_0a^2}{M_\ast}\right)^{-1}\left(\frac{c_s}{a\Omega_p}\right)^4\Omega_p^{-1}\label{eq:58}\\
  &= \left( \frac{2}{\pi} \right)^{\frac{1}{2}} \frac{\gamma \left( \gamma - 1 \right) H }{\lambda} \tau_{\text{thermal}},\label{eq:59}
\end{align}
where $\tau_\mathrm{thermal}$ is given by Eq.~\eqref{eq:22}.
Fig.~\ref{fig:Cambio_Elementos-Coeficientes} shows the time derivative of the orbital elements as a function of the luminosity for the runs of Fig.~\ref{fig:Cambio_Elementos} and runs performed with other values of the initial eccentricity or inclination. The time derivative has been obtained by measuring the difference in the values of the orbital element at $t=5$~orbits and $t=1$~orbit, i.e. at times separated by an integer number of orbital periods, in order to get rid of the oscillations seen in Fig.~\ref{fig:Cambio_Elementos}.
We see in Fig.~\ref{fig:Cambio_Elementos-Coeficientes} that our results show a good agreement with linear theory, except for the precession rates (either of the periapsis or the line of node, depending on the experiment) measured for the lowest value of the eccentricity or inclination. The slope of the dependence measured in the numerical experiments is marginally smaller than that expected from theory, in line with our findings of section~\ref{sec:corot-offs-x_p} where we measured amplitudes typically $20$~\% smaller than predicted by theory. We also see in this figure that the damping rates of eccentricity and inclination for a non-luminous planet embedded in a disc with thermal diffusion (the yellow symbols at $L/L_c=0$) are \emph{much} larger than the damping rates in an adiabatic disc (by a factor $7$ for the eccentricity and $6$ for the inclination). This is the first numerical confirmation that these damping rates on a non-luminous planet are considerable. \citet{2017arXiv170401931E} had mentioned that in discs with radiative thermal diffusion, non-luminous planets seemed to have their eccentricity and inclination damped more vigorously than in isothermal discs (their section~3.2), but their resolution was low, and the effect was not analysed. Subsequently \citet{2019MNRAS.485.5035F} found that the eccentricity and inclination of a non-luminous planet should be damped at a rate much larger than that predicted for isothermal or adiabatic discs, along the lines of Eqs.~\eqref{eq:54} and~\eqref{eq:55} but their prediction had not been supported by numerical simulations so far.

We further comment that as the planet evolves freely in the disc, it can also migrate, in principle. The migration rate arising from thermal interactions only can be written, using Eq.~(\ref{eq:22}) as well as Eq.~(146) of \citet{2017MNRAS.472.4204M}, under the form: $\dot{a}/a = 1.288 \pi \eta h^{2} \left( \ell - 1 \right) / \tau_{ \textup{thermal}}$. Comparison of this expression to the first term of the right hand side of Eqs.~(\ref{eq:54}) and~(\ref{eq:55}) shows that the evolution of the semi-major axis under the action of thermal forces occurs on much longer timescales than that of eccentricity and inclination (by a factor $\sim h^{-2}$). It is therefore perfectly safe to neglect migration over the short duration of our runs. Furthermore, note that in the runs presented here, we have $\eta=0$, so that the thermal torque does not contribute to migration and the planet is only subjected to the resonant torque, which induces even longer migration timescales.

\section{Discussion}
\label{sec:discusion}
We discuss hereafter some extensions or consequences of the analysis presented above.
\subsection{Regime of larger eccentricity or inclination}
\label{sec:regime-larg-eccentr}
\begin{figure}
  \centering
  \includegraphics[width=\columnwidth]{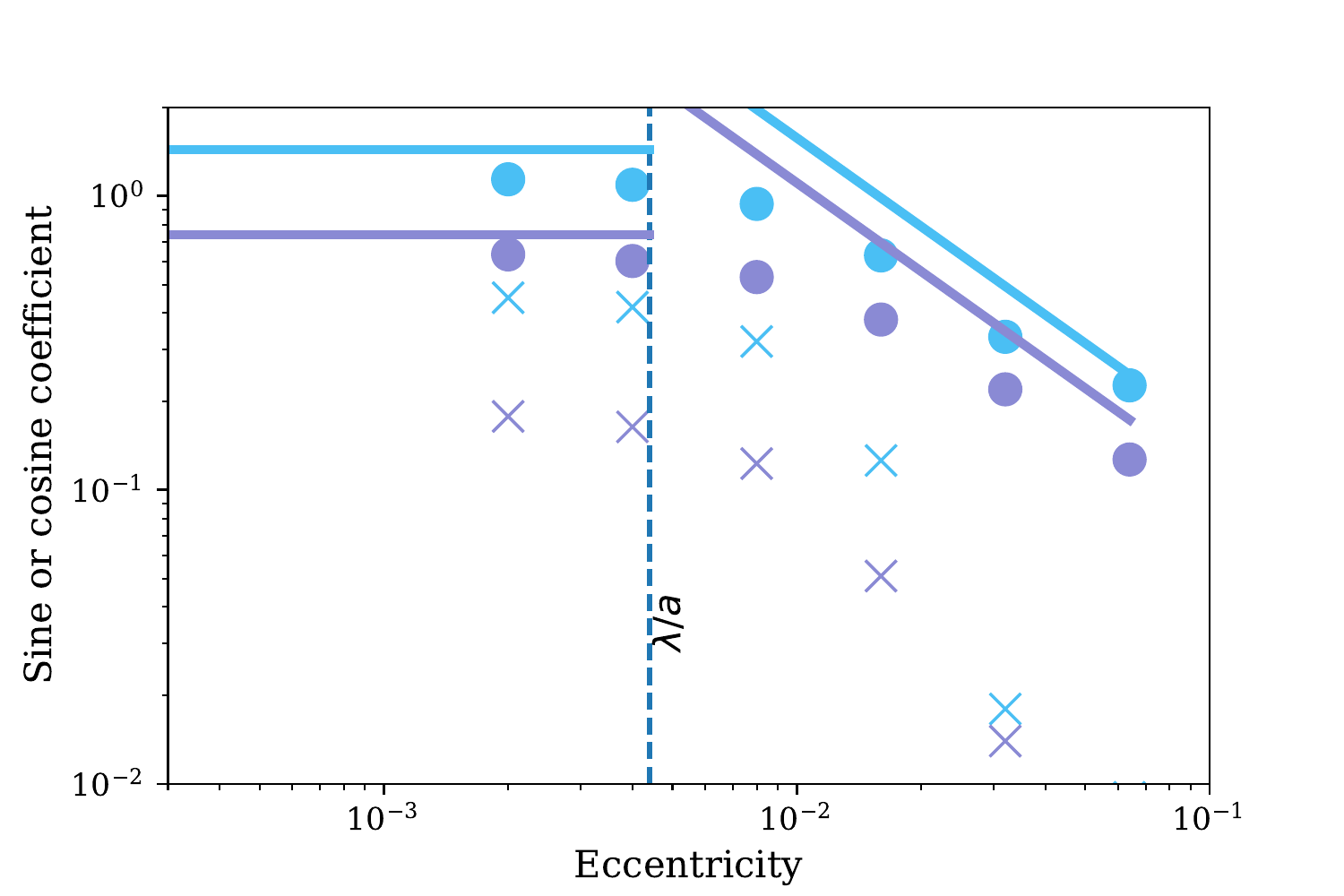}
  \caption{Absolute value of the cosine and sine coefficients of the horizontal components of the heating force, as a function of the eccentricity. Values relative to $F_x$ are represented in blue (in the electronic version of this manuscript) while values relative to $F_y$ are represented in purple. The horizontal lines show the theoretical values of $f_x^s$ and $f_y^c$ at low eccentricity, given by Eqs.~(\ref{eq:7}) and~(\ref{eq:8}). The tilted solid lines show the trends given in Eq.~(\ref{eq:69}). The disc symbols show the values measured for $f_x^s$ and $f_y^c$ for the runs described in the text. The cross symbols show the values measured for $f_x^c$ and $f_y^s$ for these same runs. For the largest value of the eccentricity, they have values below the minimum of the figure. All coefficients are positive, except $f_x^c$ which are all negative. The vertical dashed line shows the value of the eccentricity for which the radial excursion is equal to the thermal lengthscale.}
  \label{fig:eklund}
\end{figure}

The analysis presented here considers epicyclic and vertical excursions smaller than the thermal lengthscale, itself in general a small fraction of the disc scale height. Under these circumstances, the damping or growth of the eccentricity or inclination are exponential, and these two orbital parameters vary independently. When this hypothesis breaks down, the evolution of the eccentricity and inclination enters another regime \citep{2017arXiv170401931E,2017A&A...606A.114C}, in which the thermal perturbation takes the form of a ``cometary'' trail, which can be either hot or cold (with respect to the adiabatic case) depending on whether the planet's luminosity is super- or sub-critical. In this regime, the growth of the eccentricity or inclination of planets with super-critical luminosity is not exponential with time and these two parameters no longer evolve independently \citep{2017arXiv170401931E}. We can work out, in this regime, the behaviour of the sine and cosine coefficients of the different components of the force. We firstly consider an eccentric, non-inclined planet with $\lambda \ll ea \ll H$. The planet's velocity in the frame of the gas is $\left(ae\Omega_p\sin\Omega_pt,\frac 12ae\Omega_p\cos\Omega_pt,0\right)^T$. The planet is in the headwind regime, hence the thermal force has the value given by Eq.~(\ref{eq:43}), so that the horizontal components of the force exerted on the planet have the expression:
\begin{eqnarray}
  F_x&=&\frac{\gamma(\gamma-1)GmL}{2\chi c_s^2}S(\Omega_pt)\label{eq:60}\\
  F_y&=&\frac{\gamma(\gamma-1)GmL}{2\chi c_s^2}C(\Omega_pt)\label{eq:64},
\end{eqnarray}
where $S(x)\equiv \sin(x)/(\sin^2x+\cos^2x/4)^{1/2}$ and $C(x)\equiv \cos(x)/2(\sin^2x+\cos^2x/4)^{1/2}$. Using Eqs.~(\ref{eq:1}) and~(\ref{eq:12}), one can recast Eqs.~(\ref{eq:60}) and~(\ref{eq:64}) respectively as:
\begin{equation}
  F_x=\frac{\pi\lambda}{ea}S(\Omega_pt)\label{eq:65}
\end{equation}
and
\begin{equation}
  F_y=\frac{\pi\lambda}{ea}C(\Omega_pt).\label{eq:66}
\end{equation}
The fundamental mode of $S(\Omega_pt)$ has no cosine components and reads $1.141\sin\Omega_pt$, while that of $C(\Omega_pt)$ has no sine component and is $0.803\cos\Omega_pt$. We therefore infer that in this regime we have:
\begin{equation}
  f_{x}^{c} = 0.0;
  \hspace{0.15cm}
  f_{x}^{s} = + 1.141\pi\lambda/(ae);
    \hspace{0.15cm}
    f_{y}^{c} = +0.803\pi\lambda/(ae); \hspace{0.15cm} f_{y}^{s} = 0.\label{eq:69}
\end{equation}
We have undertaken an additional set of calculations with a luminous planet ($L=2L_c$), with an eccentricity $e=2^j\times 10^{-3}$, for $j\in [1,6]$. We plot in Fig.~\ref{fig:eklund} the cosine and sine coefficients of the force measured in these calculations. 
We see that the coefficients roughly decay inversely proportionally to the eccentricity when the radial excursion exceeds the thermal lengthscale, as expected from Eq.~(\ref{eq:69}), whereas for lower values of the eccentricity they have a value compatible with that given by Eqs.~(\ref{eq:7}) and~(\ref{eq:8}). We also see that the coefficients $f_x^c$ and $f_y^s$ decay sharply in the headwind regime, and become more than one order of magnitude smaller than $f_x^s$ and $f_y^c$ for the largest values of the eccentricity considered here, in line with the expectation that these coefficients should be null in the headwind regime, as seen in Eq.~(\ref{eq:69}). We note, as in previous sections, that the coefficients obtained from numerical simulations are below the theoretical expectations. There are several reasons for this discrepancy: (i) the hot trail is under-resolved,  increasingly more so at larger eccentricity, since its lengthscale is inversely proportional to the planet's velocity; (ii) the lengthscale separation between the thermal lengthscale, the planet's radial excursion and the pressure scaleheight is imperfect, as the ratio $H/\lambda$ is not large enough to accommodate a radial excursion much larger than $\lambda$, and much smaller than $H$.

Substituting the values of force coefficient of Eq.~(\ref{eq:69}) in Eq.~(\ref{eq:18}), we obtain:
\begin{equation}
  \langle\frac{de}{dt}\rangle=1.37\frac{\ell-1}{\tau_\mathrm{thermal}}\frac{\pi\lambda}{a}.\label{eq:67}
\end{equation}
The eccentricity no longer has a growth or decay that is exponential in time in the headwind regime, but rather linear with time. If one considers only the heating part in Eq.~(\ref{eq:67}) (i.e. the term in $\ell$ only), then using Eqs.~(\ref{eq:13}), (\ref{eq:22}), (\ref{eq:23}) and~(\ref{eq:43}), Eq.~(\ref{eq:67}) can be recast as:
\begin{equation}
  \langle\frac{de}{dt}\rangle=1.37\frac{F_\mathrm{dyn}}{m\Omega_pa},
\end{equation}
which is the same as Eq.~(21) of \citet{2017arXiv170401931E}. We have also performed a similar study for large values of the inclination (not shown here), and found similar results. The force coefficients expected in that case are $f_z^s0=$ and $f_z^c=4\lambda/(ia)$.

We finally comment that in a forthcoming work (Cornejo et al., submitted), we present an implementation of the heating forces in an N-body code that includes the different regimes studied here.

\subsection{Behaviour with the planetary mass}
\label{sec:behav-with-plan}
Our analysis assumes that the mass of the planet is sufficiently small that the thermal forces it is subjected to are those given by linear theory. In the regime of dynamical friction, \citet{2020MNRAS.495.2063V} have assessed the planet mass beyond which linear theory ceases to be valid, and found it to be of order of
\begin{equation}
  \label{eq:critmass}
  M_c\equiv \chi c_s/G.
\end{equation}
They found that the transition from the linear regime to a regime in which the force is a progressively smaller fraction of that given by linear theory occurs around $\sim 2M_c$ for the cold force, and around $\sim 4M_c$ for the heating force. While this behaviour ought to occur, in our context, for a large corotation offset, since in these conditions the force exerted on the planet is in the regime of dynamical friction (see section~\ref{sec:large-corot-offs}), it is not clear whether this is the case for a small corotation offset. Numerical simulations of thermal forces on a planet on a  circular orbit \citep{2021MNRAS.501...24C} suggest that the decay of the force at larger mass is similar to the decay of dynamical friction. It is therefore reasonable to assume a similar behaviour occurs at small corotation offset, as the slightly time varying thermal disturbance resembles that of the circular case. Nevertheless, this statement should be confirmed by numerical simulations with a very high resolution, able to resolve the Bondi sphere of the planet \citep{2020MNRAS.495.2063V}.

\subsection{Regime of low diffusivity and maximum damping rate}
\label{sec:regime-low-diff}
Consider an inclined, cold planet in a disc with a given corotation offset. If the thermal diffusivity decreases, so do the thermal lengthscale (see eq.~\ref{eq:1}) and the damping time of inclination (see eq.~\eqref{eq:22}: the damping is increasingly stronger. However, at a certain point, the thermal lengthscale becomes smaller than the corotation offset. When that happens, the planet is in the regime of large corotation offset and the coefficient $f_z^c$ in Eq.~\eqref{eq:19} that provides the inclination damping rate is no longer the constant of Eq.~\eqref{eq:9}, but a function $\propto \lambda/x_p$, given by Eq.~\eqref{eq:47}.  The occurrences of $\lambda$ in Eq.~\eqref{eq:22} then cancel out and the damping time becomes constant. The shortest damping time by the cold force attainable is therefore comparable to that given by Eq.~\eqref{eq:22}, in which we substitute $\lambda$ with $x_p$. It is possible to evaluate precisely the ratio of the damping timescale  of inclination $\tau_i^{[\mathrm{cold}]}$ by the cold force in the regime of large corotation offset to the damping timescale by resonant wave-launching $\tau_i^{[R]}$ of \citet{2004ApJ...602..388T}. Using Eqs.~\eqref{eq:1}, \eqref{eq:11}, \eqref{eq:13}, \eqref{eq:19} and~\eqref{eq:47}, we obtain for the former:
\begin{equation}
  \label{eq:61}
  \tau_i^{[\mathrm{cold}]} = \frac{3}{2\pi}\frac{c_s^2\Omega_px_p}{(\gamma-1)G^2m\rho_0},
\end{equation}
while the latter is simply:
\begin{equation}
  \label{eq:62}
  \tau_i^{[R]}=\frac{\tau_\mathrm{wave}}{0.544}.
\end{equation}
The ratio of these two quantities is therefore:
\begin{equation}
  \label{eq:63}
  \frac{\tau_i^{[\mathrm{cold}]}}{\tau_i^{[R]}}=\frac{3\times 0.544}{\sqrt{2\pi}\gamma(\gamma-1)}h\eta \approx 1.16h\eta.
\end{equation}
As mentioned above, this ratio is independent of the thermal diffusivity (provided this quantity is small enough that the corotation offset is larger than the thermal lengthscale).  With $\eta\sim 1$ and $h\sim 0.05$, the damping time of the inclination by the cold force can be twenty times shorter than the damping time obtained from wave-launching at the disc's resonances. While we have not worked out in section~\ref{sec:large-corot-offs} the coefficients of the horizontal force, the plots of Fig.~\ref{fig:Coeficientes-NoTanLineal} suggest a very similar behaviour, so that the shortest eccentricity damping time by the cold force should be comparable to that of the inclination, implying again a damping rate $\sim h^{-1}$ larger than that due to resonant wave-launching.

\subsection{Evaluating thermal forces in practice}
\label{sec:evaltherforcesinprac}
\begin{figure}
  \centering
  \includegraphics[width=\columnwidth]{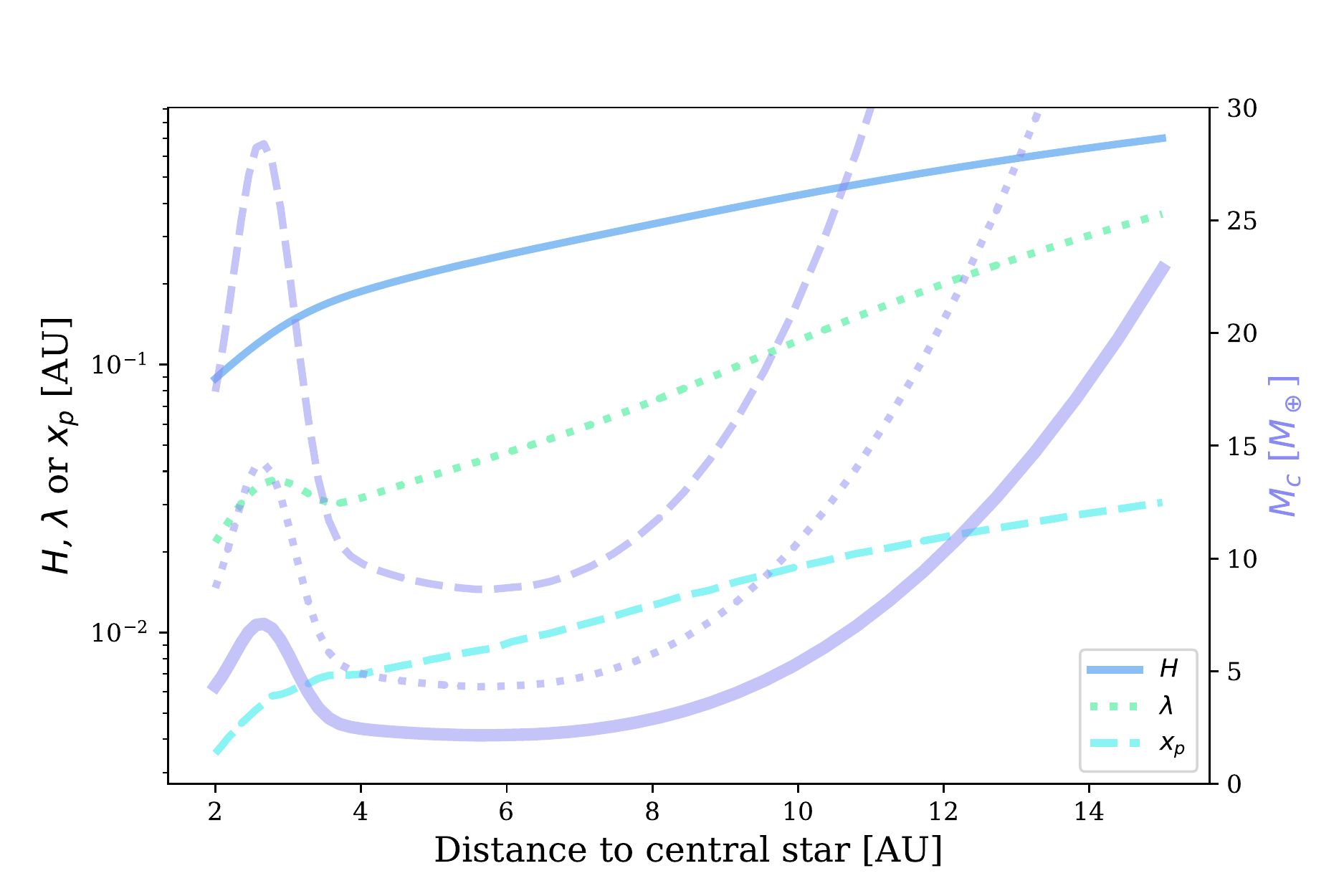}
  \caption{Pressure scale height $H$, thermal lengthscale $\lambda$
    and corotation offset $x_p$ (left axis) for the \textit{Standard
      Accretion Disc} of \citet{2015MNRAS.452.1717L}. The solid thick
    line (in purple in the electronic version of this manuscript)
    shows the critical mass $M_c$ (right axis) of
    Eq.~(\ref{eq:critmass}), while the dotted and dashed lines with
    same colour show respectively $2M_c$ and $4M_c$: the former
    corresponds to the mass at which the cold force has approximately
    half the value given by linear theory, whereas the latter
    correspond to the mass at which the heating force has
    approximately half the value given by linear theory. We see that
    the thermal lengthscale $\lambda$ is always much larger than the
    corotation offset, and that the pressure lengthscale is likewise
    always larger than the thermal lengthscale over the radial range
    considered. The maximum value of the ratio of these quantities is
    $\sim 6$, and is reached at $5-6$~AU. The minimal critical mass is
    reached at comparable radii, and is $\sim 2\;M_\oplus$. The model
    data was kindly provided by Elena Lega.}
  \label{fig:scaleslega}
\end{figure}

In addition to the usual quantities used to evaluate the resonant force between the planet and the disc (the star's and planet's masses, their separation, as well as the disc's surface density and aspect ratio), the evaluation of thermal forces requires the knowledge of the disc's thermal diffusivity $\chi$, and, for the heating force, of the planet's luminosity. The thermal diffusivity can be evaluated from the disc's temperature, density and opacity \citep[see e.g. equation~34 of][as well as references \textit{therein}]{2017MNRAS.471.4917J}. In turn, it can be used to evaluate the thermal lengthscale given by Eq.~(\ref{eq:1}) and the critical mass beyond which thermal effects start to decay, given by Eq.~(\ref{eq:critmass}). It is instructive to assess the magnitude of these quantities in the planet forming regions of a typical model of protoplanetary disc, and to compare the thermal lengthscale to the pressure scale height and corotation offset. For this purpose, we use the \textit{Standard Accretion Disc} model of \cite{2015MNRAS.452.1717L}. The results are depicted in Fig.~\ref{fig:scaleslega}.
This figure shows that the corotation offset is a minute fraction of the thermal lengthscale over the whole radial range considered, so that one can simply use the coefficients of Eqs.~(\ref{eq:7})--(\ref{eq:9}) to evaluate the thermal force, provided the eccentricity and inclinations are sufficiently small. The somehow extreme case considered in sections~\ref{sec:large-corot-offs} and~\ref{sec:regime-low-diff} of a sizeable corotation offset may concern specific locations where the thermal diffusivity (hence the thermal lengthscale) is significantly smaller than in the rest of the disc, for instance because of a larger opacity, as may happen in a dusty ring. Similarly, Fig.~\ref{fig:scaleslega} also shows that the pressure scaleheight is significantly larger than the thermal lengthscale over the radial range considered. We finally comment that the thermal diffusivity evolves as the disc ages. \citet{2017MNRAS.465.3175M} find, using template disc models of \citet{2015A&A...575A..28B}, that it tends to decrease, by more than an order of magnitude, from $t=300$~kyr to $t=1$~Myr. This implies that the critical masses decreases by the same factor, and that thermal forces become increasingly important for planets with sub-critical mass.

\section{Conclusions}
\label{sec:conclusiones}
We have performed a large number of high resolution, three-dimensional calculations of a low-mass planet with a small eccentricity or inclination, embedded in a disc with thermal diffusion. We have entertained the case of non-luminous planets, which perturb their surroundings exclusively through their gravitational potential, and the case of luminous planets, which, in addition, inject energy into the neighbouring gas at a constant rate, leading to a radiative feedback on the force from the disc. We have obtained the time varying force on the planet arising from thermal effects, and compared it to predictions from linear theory. With the resolution that we have adopted (of 6 or 8 cells per thermal lengthscale), we find a reasonable agreement on the amplitude of the oscillations of the three components of the force (which is in general $20$~\% smaller in numerical simulations than predicted by theory), and an excellent agreement on the phase, with an r.m.s. of the phase difference between all our numerical experiments and theory of $0.1$~rad. Our results also confirm that at low planetary luminosity, the action of thermal effects is a strong damping of eccentricity and inclination, whereas this behaviour is reversed, so that the eccentricity and inclination grow, when the luminosity exceeds a threshold found to be in simulations very close to the critical luminosity $L_c$ given by Eq.~\eqref{eq:13}. This watershed luminosity has exact same expression as the critical luminosity at which the dynamical friction arising from thermal effects reverses from drag to thrust \citep{2019MNRAS.483.4383V,2020MNRAS.495.2063V}, and also exact same expression as the critical luminosity at which the thermal torque exerted on a planet on a circular orbit changes sign \citep[thereby reverting migration, ][]{2021MNRAS.501...24C}.

An important result of this work is the confirmation that a non-luminous planet undergoes a much stronger damping of its eccentricity and inclination when embedded in a disc with thermal diffusion than in an adiabatic or isothermal disc. The damping time can be typically one order of magnitude shorter than that due to resonant wave-launching. While a correct description of the interaction of low-mass protoplanets with the disc should include the radiative feed back arising from their luminosity, N-body models that dismiss this aspect of the dynamics but include the gravitational interaction with the gas should at the very least implement the damping of eccentricity and inclination by the cold force, simply because real protoplanetary discs do experience (radiative) heat diffusion, and it is the mere occurrence of thermal diffusion that alters so dramatically the damping rates. These effects are so strong that they dwarf those due to the resonant interaction with the disc, to the point that the latter is virtually irrelevant for low-mass planets.

\section*{Acknowledgements}

Computational resources were available thanks to a Marcos Moshinsky Chair and to CONACyT's grant 178377. S.C. acknowledges a scholarship from CONACyT, M{\'e}xico. F.M. gratefully acknowledges support from grants UNAM-DGAPA-PASPA and UNAM-DGAPA-PAPIIT IG-101-620, and the University of Nice-Sophia Antipolis and the Laboratoire Lagrange at the Observatoire de la C\^ote d'Azur for hospitality. The work of R.O.C was supported by the Czech Science Foundation (grant 21-23067M) and by a postdoctoral CONACyT grant. S.F. acknowledges support from grant UNAM-DGAPA-PAPIIT IA103421.

\section*{Data Availability}

The FARGO3D code is publicly available at \href{https://bitbucket.org/fargo3d/public}{this address}. The setups specifically developed for the study presented here can be obtained from the corresponding author upon reasonable request.



\bibliographystyle{mnras}



\appendix

\section{Convergence study}
\label{Convergencia}
We analyse the effect of mesh resolution on thermal forces by varying the number of zones in all directions while maintaining the size of the box. From one experiment to the next, we vary the number of zones by a factor $\sqrt{2}$. Table~\ref{tab:convergencia} shows in detail the number of zones we used in each of the experiments. 

\begin{figure*}
  \begin{center}
    \includegraphics[width=.85\textwidth]{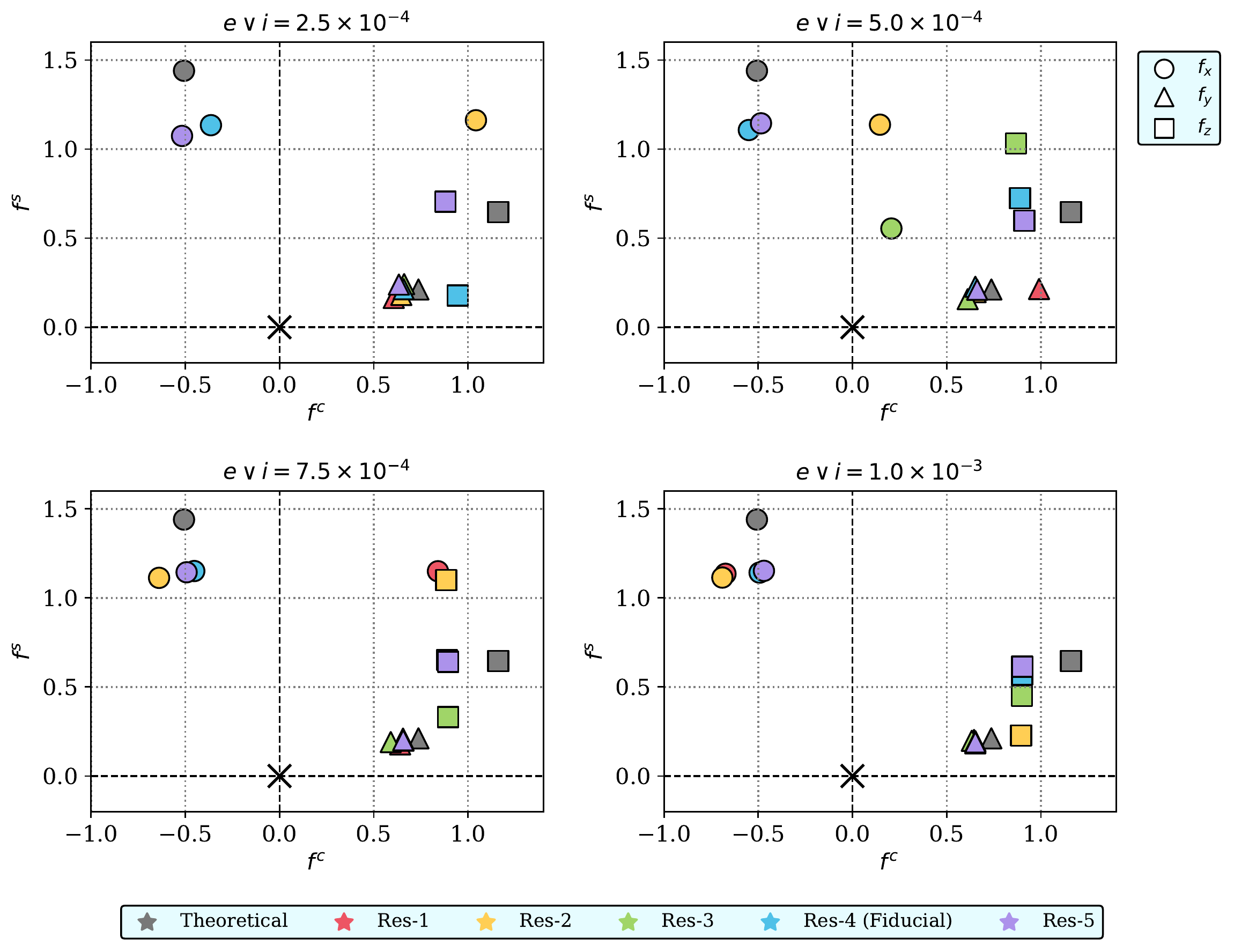}
  \end{center}
  \caption{This figure shows the individual force coefficients: the circles represent the coefficients $f_{x}$, the triangles the coefficients $f_{y}$, and the squares the coefficients $f_{z}$. The colour indicates the resolution of the runs with which a coefficient has been obtained (see Tab.~\ref{tab:convergencia}). The hot runs for this figure have all been performed with $L=2L_c$.}
    \label{fig:Convergencia-Nube}
\end{figure*}

The normalised cosine and sine coefficients calculated with the different resolutions are shown in figure~\ref{fig:Convergencia-Nube}. We see that resolutions~1 to~3 have significant outliers, while the only outlier at resolution~4 (the resolution used throughout this work) is the one already seen in Fig.~\ref{fig:Nube}, corresponding to the vertical component of the force for $i=2.5\times 10^{-4}$. Increasing the resolution further (i.e. at resolution~5) only brings minor improvements, if any, except for the outlier of resolution~4. Given the impractical cost of simulations at resolution~5 on the platform on which this study was undertaken, we have opted for resolution~4.

\begin{table}
	\centering
	\caption{Number of zones used for the setups with eccentric and inclined planets, respectively.}
	\label{tab:convergencia}
\begin{tabular}{llccc}
\hline
    Name   & Description & $N_{x}$ & $N_{y}$ & $N_{z}$ \\
\hline
\hline
\multicolumn{5}{c}{\begin{tabular}[c]{@{}c@{}}Setup for eccentric planet (half disc)\end{tabular}} \\
\hline     
    Res-1 & $\times 2^{-3/2}$  &  321  &  288 &  48  \\
    Res-2 & $\times 2^{-2/2}$ &  454  &  408 &  68 \\
    Res-3 & $\times 2^{-1/2}$  &  642  &  577 &  96 \\
    \textbf{Res-4} & \textbf{Fiducial}  &  908  &  816 & 136 \\
    Res-5 & $\times 2^{+1/2}$  & 1284  & 1154 & 192  \\
\hline
\multicolumn{5}{c}{\begin{tabular}[c]{@{}c@{}}Setup for inclined planet (entire disc)\end{tabular}} \\
\hline     
    Res-1 & $\times 2^{-3/2}$  &  251 & 228 &  76 \\
    Res-2 & $\times 2^{-2/2}$ &  354 & 323 & 107 \\
    Res-3 & $\times 2^{-1/2}$  &  501 & 457 & 151 \\
    \textbf{Res-4} & \textbf{Fiducial}  &  709 & 646 & 214  \\
    Res-5 & $\times 2^{+1/2}$  & 1003 & 914 & 303  \\
\hline
\end{tabular}
\end{table}

\bsp	
\label{lastpage}
\end{document}